\documentclass[onecolumn,usenatbib,11pt]{mnras}
\usepackage{graphicx}
\usepackage{psfrag}
\usepackage{amsmath}
\usepackage{amssymb}
\usepackage{color,xcolor}
\usepackage{hyperref}
\hypersetup{
    colorlinks=true,
    citecolor=red,
    filecolor=black,
    linkcolor=blue,
    urlcolor=magenta,
    linktocpage=true,
    breaklinks = true
}

\usepackage{subfigure,float}
\usepackage{multirow}
\usepackage{makecell}
\usepackage[normalem]{ulem}

\usepackage{array,booktabs}
\usepackage{siunitx}
\usepackage{listings}
\usepackage[authoryear]{natbib}

\def\aj{Astron. J.}
\def\apj{Astrophys. J.}
\def\apjl{Astrophys. J.}
\def\mnras{Mon. Not. Roy. Astron. Soc.}

\def\aap{Astron. Astrophys.}
\def\apjs{Astrophys. J. Suppl.}
\def\nat{Nature}

\def \pasp {Publ. Astron. Soc. Pac}
\def \araa {Ann. Rev. Astron. Astrophys.}
\def \rmxaa {Rev. Mex. Astron. Astrofis.}

\newcommand{\beq}{\begin{equation}}
\newcommand{\eeq}{\end{equation}}

\newcommand{\ber}{\begin{eqnarray}}
\newcommand{\eer}{\end{eqnarray}}

\newcommand{\ba}{\begin{align}}
\newcommand{\ea}{\end{align}}

\def \vs {V_{\rm sys}}
\def \dv {\Delta V}
\def \dl {\Delta \lambda}
\def \nit {N_{\rm it}}
\def \l {\lambda}
\def \kmps {km/s}
\def \nb {$N_B$}
\def \fa {F_A}
\def \fb {F_B}

\def \rba {r_{B^sA}}
\def \D {\Delta}

\def \flu {$10^{-17}$ erg/cm$^{2}$/s/$\AA$}

\title[Galaxy rotation curve from cross-correlation]{A novel approach for calculating galaxy rotation curves using spaxel cross-correlation and iterative smoothing}

\author[Bag et al ]{Satadru Bag$^1$\thanks{satadru@kasi.re.kr},  Arman Shafieloo$^{1,2}$\thanks{shafieloo@kasi.re.kr}, Rory Smith$^{3}$,  Haeun Chung$^4$,
\newauthor
Eric V. Linder$^{5,6,7}$, 
Changbom Park$^{8}$, 
Y. Sultan Abylkairov$^{7}$,  Khalykbek Yelshibekov$^{7,9}$
\\
\\
$^{1}$ Korea Astronomy and Space Science Institute, Daejeon 34055, Korea\\
$^2$ University of Science and Technology, Daejeon 34113, Korea \\
$^3$ Departamento de Física, Universidad Técnica Federico Santa María, Avenida Vicuña Mackenna 3939, San Joaquín, Santiago, Chile\\
$^4$ University of Arizona, Steward Observatory, 933 N Cherry Ave, Tucson, AZ 85721, USA\\
$^5$ Berkeley Center for Cosmological Physics, University of California, 
Berkeley, CA 94720, USA\\ 
$^6$ Lawrence Berkeley National Laboratory, Berkeley, CA 94720, USA\\ 
$^7$ Energetic Cosmos Laboratory, Nazarbayev University, Nur-Sultan 010000, 
Kazakhstan\\  
$^8$ School of Physics, Korea Institute for Advanced Study, Seoul 02455, Korea\\
$^9$ Department of Physics,  University of California San Diego, La Jolla, California 92093, USA \\
}

\date{}

\pubyear{2021}

\begin{document}
\label{firstpage}
\pagerange{\pageref{firstpage}--\pageref{lastpage}}
\maketitle

\begin{abstract}
Precise measurements of the internal dynamics of galaxies have proven of great importance for understanding the internal dark matter distribution of
galaxies. We present a novel method for measuring the line-of-sight (LOS) velocities across the face of galaxies by cross-correlation of spectral pixels (spaxels) and an iterative method of smoothing. 
On simulated data the method can accurately recover the input LOS velocities for different types of spectra (absorption line dominated, emission line dominated, and differing shapes of the continuum), and can handle stellar population radial gradients. Most important of all, it continues to provide reliable measurements of LOS velocities with reasonable uncertainties even when the spectra are very low signal-to-noise (approaching $\sim 1$), which is a challenge for traditional template-fitting approaches. 
We apply our method to data from a real MaNGA galaxy as a demonstration and
find promising results with good precision.
This novel approach can be complementary to existing methods primarily based on template fitting. 
\end{abstract}


\section{Introduction} 
The importance of measuring the line-of-sight velocities in astronomy cannot be overstated since they shed light on several key questions in cosmology and astrophysics pertaining to structure formation, galaxy dynamics in clusters (that first revealed the existence of dark matter halos \citep{Zwicky1933}), distance measures, as well as internal galaxy (and globular cluster) dynamics.  In particular, the internal dynamics of galaxies has drawn major amount of interests for decades as this is a fundamental quantity, a part of the empirical scaling properties that almost all galaxies follow, such as the Tully-Fisher (Luminosity vs rotational velocity) \citep{TullyFisher}, Faber-Jackson (Luminosity  vs velocity dispersion) \citep{FaberJackson}, Fundamental plane (Luminosity  vs $R_{\rm eff}$ vs velocity dispersion) \citep{Gudehus,Cole1994}. Furthermore, if baryons are in virial equilibrium with the galaxy’s total potential, the dynamics can be used to measure the total mass of the system (the dynamical mass) and the mass-distribution \citep{Rubin1978, Bosma1981, vanAlbada1985,2011MNRAS.415..545T,2020MNRAS.496.1857L,2021MNRAS.503.5238K}; see \cite{Sofue2001} for a review. This is what historically first revealed that dark matter dominates in the outskirts of galaxies \citep{Rubin1970, Roberts1973, Rubin1980}. One can also measure the dynamics of baryons that are not in equilibrium, e.g., gas outflows \citep{1988ASSL..142..285R, 2010arXiv1001.2480S, 2016A&A...590A.125C, 2021A&A...654A.128C}. Therefore, the precise measurement of the internal dynamics of galaxies remains as one of the most important aspects of studying galaxy evolution. 

Line-of-sight velocities of baryons are typically measured by Doppler shift of lines in the spectra. For atomic gas the primary line of interest was the 21cm line emission \citep{Ewen,2006ApJ...640..751C,2008AJ....136.2648D,2020ApJ...889...10D}. For stellar components, and ionised gas, one can consider their spectral lines (H$\alpha$ and NII are especially bright lines in the optical part of the spectrum). Traditionally this was done with long slit spectroscopy, resulting in a 1D rotation curve \citep{1992A&AS...94..175A,1993ApJ...415L..95V,2001ARA&A..39..137S,2004A&A...416..475M,2011MNRAS.416.1936T,2015MNRAS.451.1004E}. Spectra also provides valuable information on other properties of stars (other stellar dynamics (e.g., dispersion), stellar populations, ages, metallicities, alpha abundances). Recently, IFUs have become very important. Here spectra are measured in a 2D grid of spaxels across the face of the galaxy. Many large area surveys, e.g.\ CALIFA (\href{https://califa.caha.es/}{https://califa.caha.es/})  \citep{2012A&A...538A...8S,Califa}, MANGA (\href{https://www.sdss.org/surveys/manga/}{https://www.sdss.org/surveys/manga/}) \citep{Manga2015,2015AJ....149...77D,2016AJ....152..197Y,2017AJ....154...86W}, SAMI  (\href{http://sami-survey.org/} {http://sami-survey.org/}) \citep{SAMI,Croom2021}, and in the future Hector \citep{Hector}, have databases of thousands of galaxies.

Traditionally matching the spectra is done with template fitting \citep{Cappellari2004,Fernandes2005,Ocvirk2006,Walcher:2006hd,Koleva:2009kt,Sanchez,2017MNRAS.466..798C}. We get a lot of information from the fit - dynamics, stellar populations, etc \citep{2012IAUS..284...42A,2017MNRAS.466..798C,2019A&A...622A.103B,2021ApJS..254...22J}. But results are strongly influenced by assumptions built into the templates. Also, when S/N is low, even when it is still possible to measure stellar populations, it becomes difficult to measure line-of-sight velocities with the template fitting approach. S/N can become low in low surface brightness galaxies, or in the outskirts of any galaxy. To overcome this, IFU studies attempt to sum together the spectra from multiple spaxels to try to increase S/N using Voronoi tessellation fields \citep{voronoi,Cappellari2003,2015A&A...573A..59G,2019MNRAS.489..608F,2020MNRAS.493.3081R,2021MNRAS.507.2488G}. However, there is still a limit to how far out in radius they can go or how faint galaxies can be. For example, most galaxies in SAMI only have detected stellar dynamics out to one or two effective radii \citep{Croom2021}. If the aim is to study dark matter content, this is limiting as it is only beyond $\sim 1 R_{\rm eff}$ that the dark matter begins to dominate the potential \citep{Cappellari2013}. This in turn curtails the possibility for new discoveries from the existing data.

In this article, we seek to explore a template independent method, proposing a novel approach  to calculate the velocity differences between the pairs of spaxels. Specifically, we estimate the Doppler shifts using the cross-correlation of the respective spectra, with robust control over both regularization (smoothing) and selection of the particular parts of the spectra. 
Our method is not dependent on identifying individual features (emission/absorption lines) in the spectra, and of course independent of any template. Therefore, apart from providing a useful crosscheck on traditional methods, it can handle difficult observational conditions (e.g. variation in spectra types, low S/N, etc), as well as realistic complex non-ideal data (incomplete spectra, missing spaxels etc). We demonstrate that this approach allows success even for  lower signal to noise spectra than usual, and more robustness as it takes into account the whole spectra, not just predominantly a few  features.

The paper is organized as follows. In the next section, we explain the novel technique based on iterative smoothing and computing the cross-correlation between a pair of spectra. We validate the approach on a variety of simulations in section 3 after demonstrating the method on a pair of spaxels. In the following section, we push the method to limit by systematically testing on data with different noise levels from high to very low S/N.
In section 5, we validate the method on a MaNGA galaxy and compare the results with that from Marvin. We conclude and put useful discussion in section 6.

\section{Methodology}
\begin{figure}
\centering
\includegraphics[width=0.485\textwidth]{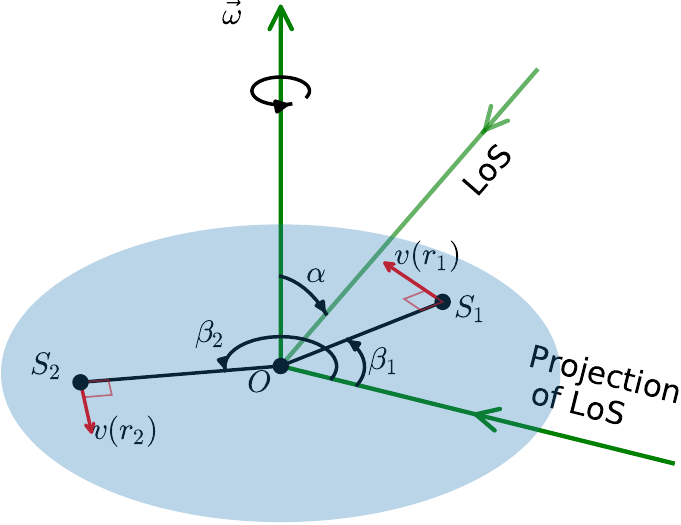}
\caption{The galactic plane is shown by the shaded surface. Suppose the line of sight (LOS) has an angle $\alpha$ with the rotational axis. The velocity components along the projection of LOS for a point on the galactic plane would be  $v(r) \sin\beta$. This can be further projected along the LOS by multiplying with the $\sin \alpha$ term.}
\label{fig:sch_angles}
\end{figure} 

For a galaxy observed with a spectrograph, we obtain spaxel (spectral pixel) for different parts of the galaxy, whether a 2D array in the case of an integral 
field spectrograph or a line from slit spectroscopy. By comparing one spaxel to another, one can measure  
wavelength shifts of spectral features in common. 

The wavelength shift from spaxel 1 to 2 is directly related to the line of 
sight velocity difference through the Doppler shift, so 
\begin{equation} \label{eq:dv}
 \Delta V= c\ \frac{\lambda_2-\lambda_1}{\lambda_1} \;.
\end{equation}
In order to relate this to the galaxy rotation curve (GRC) we must take into account 
the inclination of the galaxy with respect to the line of 
sight, with 
\begin{align}\label{eq:vels}
 V_1&=\vs + v(r_1)\sin\alpha\, \sin\beta_1\;, \\ \nonumber
 V_2&=\vs + v(r_2)\sin\alpha\, \sin\beta_2 \;,
\end{align}
where the angles are as illustrated in Figure~\ref{fig:sch_angles},  
$\vs$ is the line of sight velocity of the central spaxel, or 
in the simplest case of the whole galaxy, and $v(r)$ is the 
rotational velocity of a spaxel at distance $r$ from the 
galactic centre. 


The velocity difference between spaxels as measured by the 
wavelength shift therefore allows one to build up the galaxy 
rotation curve $v(r)$ through 
\begin{equation}\label{eq:vr}
 \Delta V=\sin \alpha\,\left[(v(r_2)\sin \beta_2-v(r_1)\sin \beta_1\right]\;, 
\end{equation}
given the values of the angles $\alpha$ and $\beta$. 
One repeats the process for many spaxel pairs, and hence 
$\Delta V$'s, to construct the GRC. 

In this paper we focus on introduction of our technique for 
measuring $\Delta V$; the extraction of $v(r)$ from those 
is well established in the literature. Our basic approach is 
to cross-correlate the spectra from spaxel pairs. 
Since observed spectra have noise, one needs to avoid 
spurious features from the noise overwhelming the true 
wavelength shift. In addition, spectra are sometimes incomplete, 
missing wavelength regions due to, for example, intervening 
atmospheric lines. 

Therefore we first smooth one spectrum of the pair while 
leaving that of the other intact; we will then also switch 
which of the pair is smoothed to compare results and prevent 
a single high noise spectrum from distorting the estimate. 
We cross-correlate the spectra pair as a function of offset 
$\delta V_{\rm off}$ for a wide range of offsets, building 
up a cross-correlation function $r(\delta V_{\rm off})$. 
The mode of this curve is expected to be the actual $\Delta V$. 

In the next subsections, we go into the technical details 
of how we carry out the smoothing, estimate the cross-correlation 
function, and then use it to obtain $\Delta V$.

\subsection{Iterative smoothing of the spectra}\label{sec:smooth} 

We smooth an observed spectrum, $F_{\rm obs}(\lambda_i)$, iteratively with a Gaussian kernel following \citet{Shafieloo:2005nd, Shafieloo:2007cs, Shafieloo:2009hi, Aghamousa:2014uya}; these papers establish the criteria for stable, accurate results. 
The smoothed spectrum in the $n$th step is obtained from the previous step as
\begin{equation}\label{eq:smooth}
 F^s_{n}(\lambda)=F^s_{n-1}(\lambda)+ \frac{1}{N(\lambda)} \sum_i\frac{\left(F_{\rm obs}(\lambda_i)-F^s_{n-1}(\lambda_i)\right)}{{\sigma_{\rm obs}}^2(\lambda_i)}\times \exp{\left[-\frac{(\lambda -\lambda_i)^2}{2 \D^2} \right]}
\end{equation}
where the normalisation term $N(\lambda)$ is given by 
\begin{equation}
 N(\lambda)=\sum_i \left(\frac{1}{{\sigma_{\rm obs}}^2(\lambda_i)}\right) \times \exp{\left[-\frac{(\lambda -\lambda_i)^2}{2 \D^2} \right]}\;.
\end{equation}
Here $F^s_{n-1}(\lambda)$ is the smoothed spectrum obtained in the previous step, i.e.\ at the $n-1$ step. We start with an initial guess for the smoothed spectrum, say $F^s_0(\l)$= constant in the first step, and continue smoothing for $\nit$ number of iterations. For sufficiently number of iteration, the final smoothed spectra should become independent of the initial guess. Note that we have two parameters in this smoothing algorithm: the smoothing scale $\D$ and the number of iterations $\nit$. These two parameters should be fixed according to the problem at hand. For example, $\D \gtrsim \delta \lambda$ where $\delta \lambda$ is the average wavelength resolution in the 
observed spectrum  and $\nit \gtrsim 5$ work well for the typical examples we consider in this work. Testing by  simulations as described in Section~\ref{sec:validsim} 
establishes that 
the final results are properly insensitive to the choice of $\D$ and $\nit$ when they are within reasonable ranges. Also note that in this method we can obtain the smoothed spectrum $F^s$ at any desired values of $\lambda$ in between the observed wavelengths $\lbrace \l_i \rbrace$. Thus this smoothing algorithm also serves the purpose of interpolation needed for cross-correlations with any arbitrary Doppler (wavelength) shift.

\subsection{Estimating $\dv$ using cross-correlation}\label{sec:cc} 
 For different relative velocities ($\dv$) we cross-correlate the spectra considering the corresponding Doppler shift in the wavelength (related through equation \eqref{eq:dv}), with one of them being smoothed 
using \eqref{eq:smooth}. 
(We also tried initial ``supersmoothing'', using a very large $\Delta_{\rm ini}$ to remove the continuum trend; this did not give any advantage over the results we present.) 
We define the weighted cross-correlation between the spectra of two spaxels (say A and B) as a function of $\dv$ as
\begin{equation}\label{eq:rab}
 r_{A^sB} (\dv)\equiv F^s_A (\l+\dl)\otimes F_B(\l) =\frac{\sum_i w_i \left[ F^s_A(\l_i+\dl)-\left<F^s_A(\l_i+\dl)\right>_w \right] \left[ F_B(\l_i)-\left<F_B(\l_i)\right>_w \right]}{\sqrt{ \sum_i w_i \left[ F^s_A(\l_i+\dl)-\left<F^s_A(\l_i+\dl)\right>_w\right]^2} ~\sqrt{\sum_i w_i  \left[ F_B(\l_i)-\left<F_B(\l_i\right>_w\right]^2 }}\;,
\end{equation}
where $\dl = \l(\dv/c)$. The index $i$ runs over the 
wavelengths sampled in the spectra. 
We employ inverse variance weighting $w_i=1/{\sigma^2_B}_i$, 
where ${\sigma_B}_i$ is the uncertainty of the second, unsmoothed 
spectrum. 
The correlation coefficient $r_{AB} (\dv)$ should have a 
maximum when $\dv$ is the same as the true velocity difference 
between the two spaxels.

\subsection{The algorithm to estimate $\dv$}\label{sec:algorithm} 

Given these ingredients, the algorithm for obtaining the velocity difference between a pair of spaxels A and B is: 
\begin{itemize}\itemsep1em 
 \item Smooth the spectra observed in one of the spaxels, say A, 
 calling the smoothed spectrum $\fa^s$.
 
 \item Cross-correlate $\fa^s$ and $\fb$ for a large number of relative velocities ($\dv$). For added robustness through crosschecks, instead of cross-correlating the whole spectra to give a single answer, 
 we divide the (unsmoothed) spectrum of B into \nb ~number of bins to utilize information from different parts of the spectrum independently. We determine the cross-correlation \eqref{eq:rab} for each bin of $\fb$ separately, i.e.\ we calculate separate 
 estimates $r_{A^sB_j}(\dv)$ for the $j$th bin where $j \in \lbrace 1, N_B \rbrace $.

 \item Find the global maxima of the correlation coefficients  $r_{A^sB_j}(\dv)$ for each bin. Let the maximum for the $j$th bin be $r^{\rm max}_{A^sB_j}$, occurring at the velocity difference $\dv_{AB_j}$. We will have \nb\ such velocity estimations corresponding to \nb\ bins. 
 
 \item Swap the two spaxels and repeat the same procedure, i.e.\  evaluate ${\rba}_j(\dv)$ and find $\dv_{BA_j}$ (always using the variance of the unsmoothed spectrum as the weight factor), 
 again separately for each of the \nb\ bins. 
 We now have $2$\nb\ number of velocity estimations. 
We expect $\dv_{BA} \approx - \dv_{AB}$ for all the bins; 
this is another crosscheck. 
 
 \item Estimate the velocity difference from these 
 $2$\nb\ estimations. As different parts of the spectra may have different signal to noise ratios, different bins can give rise to $\dv$ estimation with different degrees of accuracy. To filter 
 out noise we only accept estimates satisfying the symmetry requirement:
 the velocity estimations $|\dv_{AB}|$ and $|\dv_{BA}|$ 
 are close to each other, with   $|\dv_{AB_j}+\dv_{BA_j}| \leq 0.05\, |\dv_{AB_j}-\dv_{BA_j}|$, or $\le10$ km/s\footnote{ Note that this criterion could be set according to the problem in hand, considering several factors like the signal to noise ratio of the spectra.}.

\item From the acceptable estimates determine the final velocity 
difference $\dv_{AB}$ between the spaxels by averaging over these 
estimates. Estimate the uncertainty $\sigma_{\dv_{AB}}$ using the standard deviations.

\end{itemize}

Cross-correlating one unsmoothed spectrum with another smoothed spectrum helps to increase the effective signal to noise. Without smoothing either spectrum, the correlation coefficient curve $r(\dv)$ exhibits wiggles because of random matching of the two spectral noises. This may alter the locations of the maxima of $r(\dv)$ for different bins leading to erroneous estimations, especially for low signal to noise cases. Conversely, smoothing both spectra simultaneously leads to neglect of some uncertainties in the data, and hence affects the realistic estimation of the final $\dv$. By smoothing each spectrum in sequence in addition we can have a first, basic consistency test between $\dv_{AB}$ and $\dv_{BA}$ to make sure that we are not fitting noise and both estimations are meaningfully consistent.

\section{Validating the method using simulation} 
\label{sec:validsim}

Our aim is to test the new technique’s ability to measure the shape of a model galaxy’s rotation curve in diverse observational circumstances. Galaxies have different spectra, e.g.\ red and dead galaxies have absorption lines, while star-forming galaxies may be dominated by emission lines. And galaxies often have differing spectra as a function of radius (e.g.\ some spiral galaxies have red and dead centres and star formation in their disks). We need to ensure our method is able to successfully recover the input rotation curve even when different types of spectra are considered. Therefore, we test the method through simulations of spectra of galaxies in different scenarios, e.g.\ with absorption vs emission line dominated spectra, with different noise levels and strengths of features, radially varying spectral properties etc. 

In this first work, for simplicity we always simulate the model galaxy to be edge-on and the spaxels are placed on the major axis perpendicular to the line of sight (so $\alpha=\beta=\pi/2$). 
That is, we demonstrate the procedure in a 1D situation, with 
the full 2D galaxy data treated in a follow-up paper; this only 
affects determination of $v(r)$, not $\dv$. 
The velocity difference between two spaxels then  
simply becomes the difference in their rotational velocities, 
i.e.\ $\dv_{AB}=v(r_A)-v(r_B)$.
We assume an exponential disk for the model galaxy's stellar disk with an effective radius of $1.5$ kpc. The spaxels are distributed from $-5$ to 5 kpc, and there are 21 spaxels in total such that spaxel 11 is centred on the disk centre. For the line-of-sight dynamics of the stars in each spaxel, we assume a Polyex model \citep{Giovanelli2002,2006ApJ...640..751C}:
\beq
v_{\rm PE}(r)=v_0\left(1-e^{-r/r_{\rm PE}} \right)\left( 1+\frac{\gamma r}{r_{\rm PE}} \right) \label{eq:vrmodel} 
\eeq
where $v_0$, $r_{\rm PE}$ and $\gamma$ are free parameters that can be varied to alter the shape of the stellar dynamics at each radius. 
 When we carry out validation fits, we 
take two cases: one where the input dynamics is known (so are these parameters)
for testing, in the other it is blinded. We will find that this does not affect the fit quality.

\begin{figure}
\centering
\includegraphics[width=\textwidth]{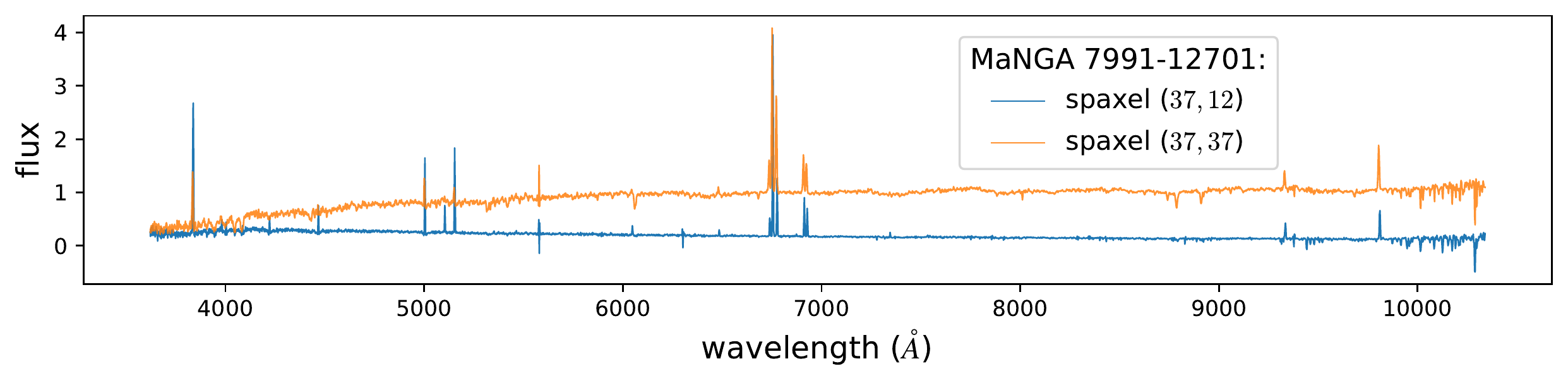}
\caption{The MaNGA galaxy 7991--12701 has emission line dominated spectra, shown here from the two spaxels (37,37) and (37,12), given 
by their coordinates in the 2D IFU grid. The flux observed in MaNGA is in the standard unit of [\flu] which has been used as the unit of flux in the relevant figures throughout the article.  The two spectra exhibit 
a similar set of features on different continua, and have a 
slight wavelength shift with respect to each other that is the manifestation of their velocity difference ($\dv \sim 168$ \kmps, 
so the 0.06\% shift is difficult to see by eye). 
The spectra in the central spaxel (37,37) has been used in the simulation for demonstrating the method in section~\ref{sec:demo} and for the test presented in section~\ref{sec:1emission}. The test demonstrated in section~\ref{sec:2emission} uses both these spectra in the simulation.
}
\label{fig:emission_spectra_7991-12701}
\end{figure}

\begin{figure}
\centering
\includegraphics[width=\textwidth]{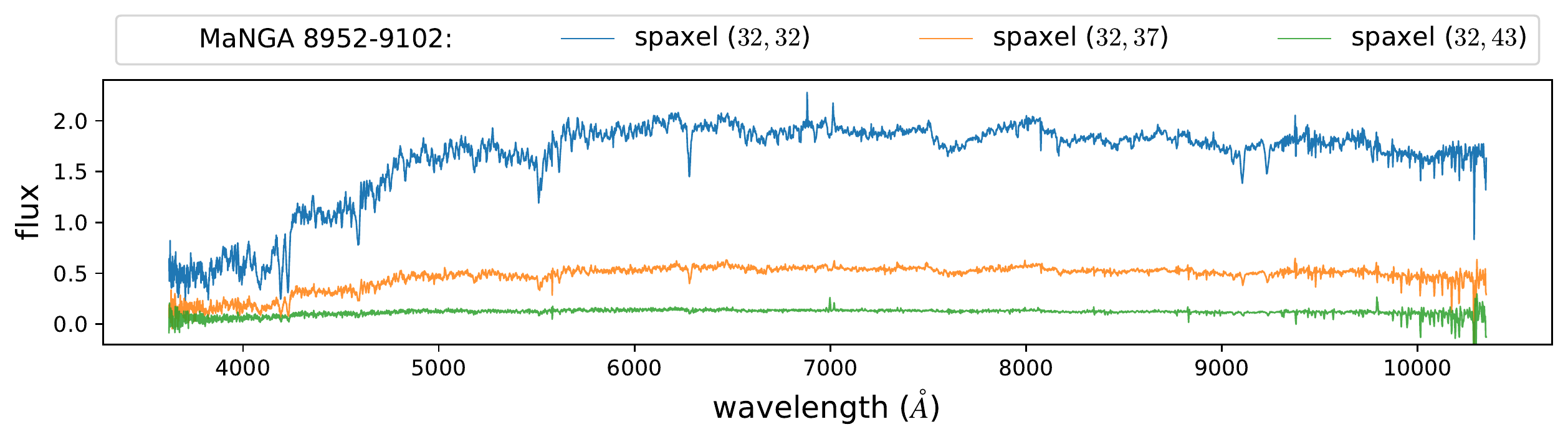}
\caption{The MaNGA galaxy 8952--9102 has absorption line or weak 
emission line dominated spectra, shown here from the three spaxels  (32,32), (32,37) and (32,43). The spectra for the central spaxel (32,32) has been used for the test demonstrated in 
section~\ref{sec:1absorption}. 
The spectrum from the spaxels (32,37) and (32,43) are used for the test in 
section~\ref{sec:absorption_vs_weak_emission}. 
} 
\label{fig:absorption_spectra_8952-9102}
\end{figure}

We now build the stellar spectrum of each spaxel. In order to test with realistic spectra, we generate the spectra in the simulations using the observed spectra of `benchmark' spaxels from two rather different MaNGA galaxies, 7991--12701 and 8952--9102\footnote{We obtain the IFU data of MaNGA galaxies from Marvin \citep{Marvin} DR15 given at \href{https://dr15.sdss.org/marvin/}{https://dr15.sdss.org/marvin/}.}. The spectra in all spaxels for the former galaxy are dominated by strong emission lines; spectra in two spaxels including the central one are shown in figure~\ref{fig:emission_spectra_7991-12701}. On the other hand, the spectra of the  MaNGA galaxy 8952--9102 show different properties in different spaxels: the central ones are absorption line dominated while the edge ones have both weak absorption and weak emission lines. Spectra from some of these spaxels are shown in 
figure~\ref{fig:absorption_spectra_8952-9102}. We then apply a Doppler shift to the spectra, according to the dynamics that the Polyex model provided. We also inject white noise into the MANGA galaxy spectra so that the signal-to-noise in each spaxel will reduce as the surface-brightness of the galaxy's disk becomes fainter exponentially in the disk outskirts, as occurs in real galaxies.

We carry out the validation procedure in steps, from initially 
assuming spectral homogeneity and noise homogeneity, then relaxing 
these conditions one by one. Our model should be considered a toy model for mocking up real IFU observations of galaxies. It allows us to test the method's ability to recover the input dynamics in a controlled manner, with a realistic set of galaxy spectra, a reasonable choice of rotation curve shapes, and with differing but controlled amounts of noise injected.

\begin{figure}
\centering
\includegraphics[width=0.8\textwidth]{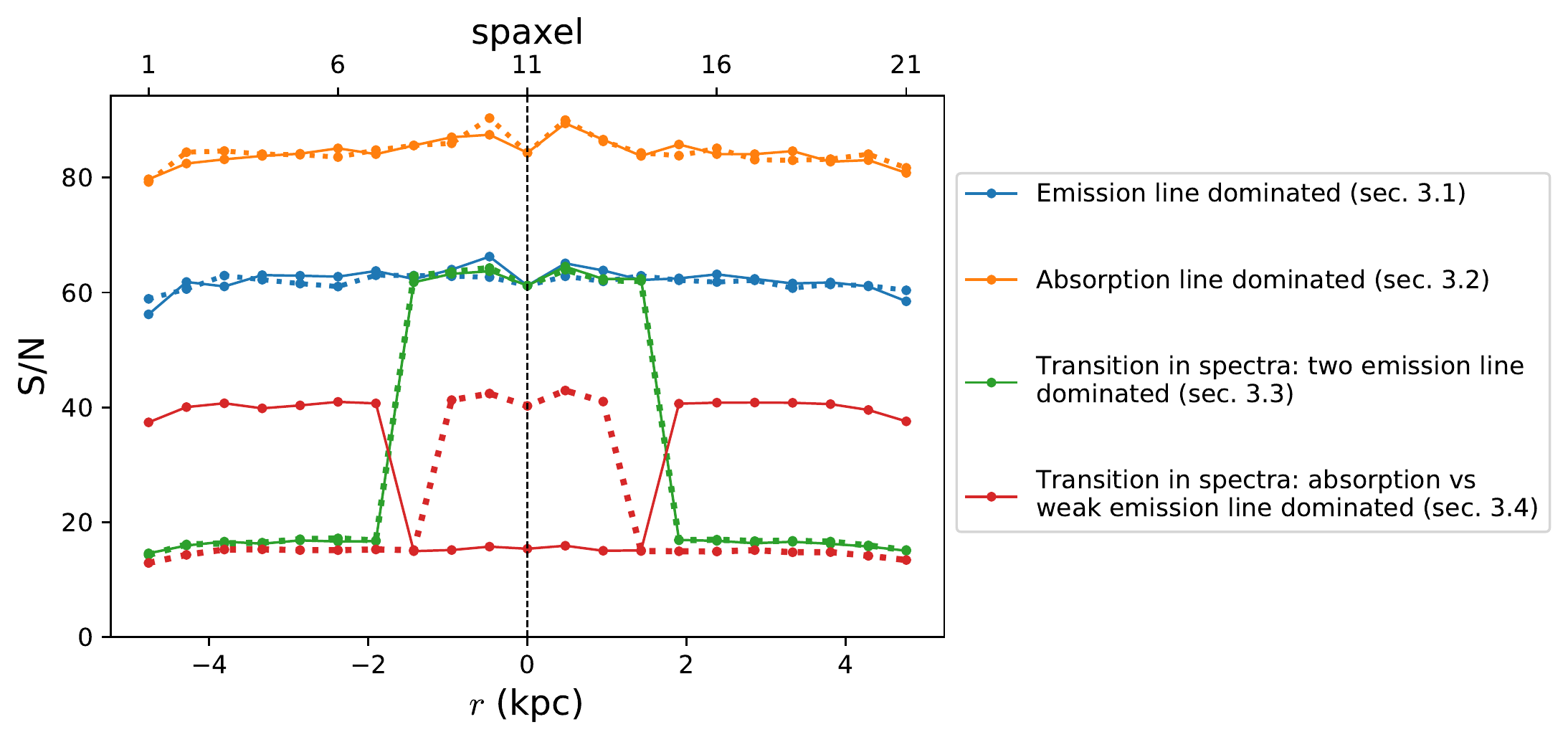}
\caption{Signal to noise ratio (S/N) for fluxes in the spaxels for different tests illustrated in 
sections~\ref{sec:1emission}--\ref{sec:absorption_vs_weak_emission}. 
The solid and dotted curves represent 
two different realizations for each test. 
Note in the mixed spectra cases 
(green and red curves) we challenge the method 
with step function transition in spectra type; 
we further challenge it in the lowest S/N case 
(red curve) by flipping the sign of the transition 
between the two realizations. 
}
\label{fig:S_N_lowNoise}
\end{figure}

We measure the signal to noise ratio (S/N hereafter) of a whole spectrum following \citet{SNR}. Figure \ref{fig:S_N_lowNoise} shows how S/N of the spectra vary across the spaxels for the different tests that we pursue in this section. The details of the tests involving a variety of different observational conditions are explained below in the respective subsections.

\subsection{Simulation with emission line spectrum} 
\label{sec:1emission}

We begin validation by simulating spectra in 21 spaxels located on the 
major axis of a galaxy, 
but fixing the properties of all the spectra in different spaxels to that of the spectrum of the central spaxel of MaNGA 7991-12701 (shown in orange in 
figure~\ref{fig:emission_spectra_7991-12701}). Each is shifted according to the line of sight velocities input in the simulation. The spectra have been observed typically within the wavelength range $\lambda=[3621.6, 10339.5]$ \AA\ with an average interval of $ \delta \lambda \approx 1.47$ \AA. However, we remove all the data points above $\lambda >9850 $ \AA\ due to the telluric contamination arising from the earth's atmosphere. 
For this first stage we also keep the noise level low and the same for all the spaxels. 
Note that even with a uniform noise level, the spectra simulated in the outer/edge spaxels would have worse signal to noise due to the outer disk of a galaxy being fainter.

\subsubsection{Determining the velocity differences} 
\label{sec:demo} 

\begin{figure}
\centering
\includegraphics[width=\textwidth]{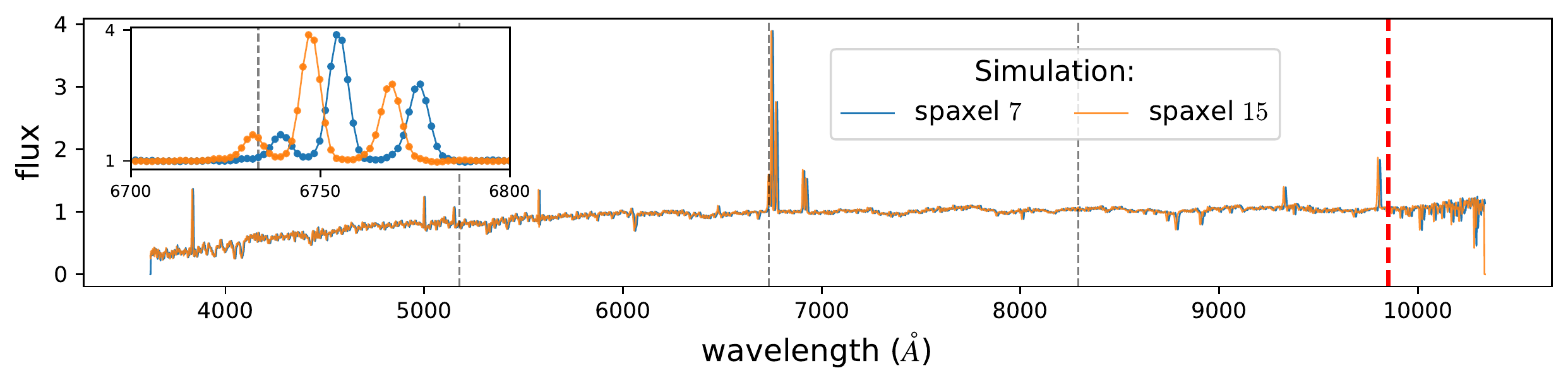}
\caption{Spectra simulated for the initial tests in 
Section~\ref{sec:1emission}, in two spaxels, 7 and 15, 
based on the spectrum observed in the central spaxel of MaNGA 7991-12701 (orange curve in 
figure~\ref{fig:emission_spectra_7991-12701}). 
The simulated spectra are shifted according to the input line of sight velocity difference (as shown in the inset which zooms into the spectral range of H$\alpha$ and NII emission lines) and sampled in equispaced 
wavelengths. 
After discarding the $\lambda > 9850$ \AA\ region due to telluric contamination, the rest of the data has been equally divided into 4 bins (each bin encloses the same number of datapoints), shown by the vertical dashed lines.
}
\label{fig:demo_spectra_cc}
\end{figure}

Since we are working with a line of simulated spaxels along the 
major axis of a galaxy, we can refer to them simply by 
their second coordinate; this also serves to distinguish simulated 
spaxels (with a single number) from MaNGA's observed spaxels 
(denoted with the 2D IFU coordinates). 
Consider simulated spectra in a pair of spaxels, $A=7$ and $B=15$, shown 
in figure~\ref{fig:demo_spectra_cc}. They are dominated by strong emission lines and derived from the same underlying spectra that is observed in the central spaxel of the MaNGA 7991-12701 galaxy. 
The true velocity difference between these spaxels is $326.83$ km/s (the corresponding Doppler shift is evident from the inset which zooms into the spectral region of H$\alpha$ and NII emission lines). 
For each of $N_B=4$ wavelength bins, we calculate the cross-correlation between one spaxel and the other 
(smoothed) one as a function of applied velocity shifts.

\begin{figure}
\centering
\subfigure[$7\bigotimes 15$]{
\includegraphics[width=0.485\textwidth]{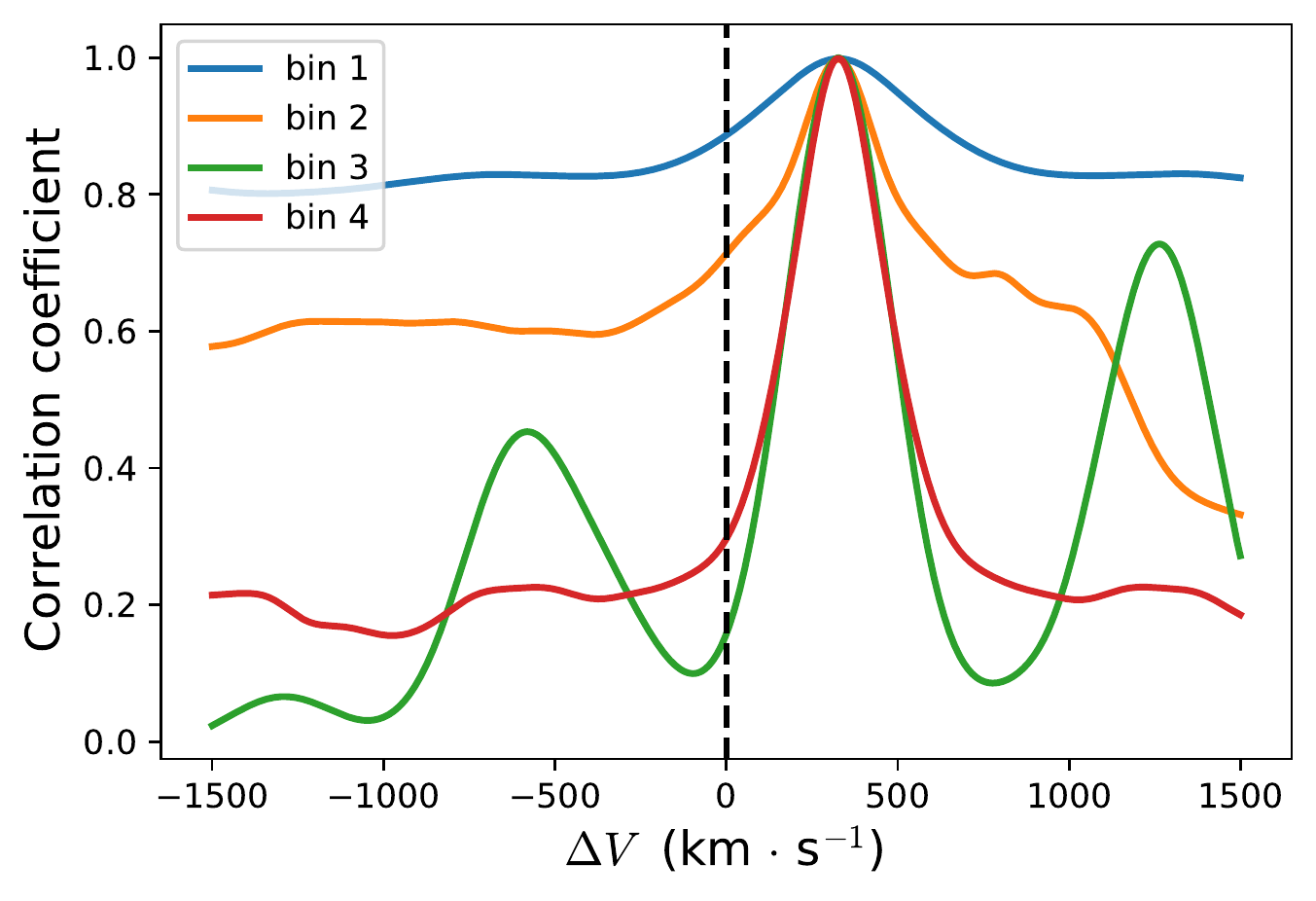}}
\subfigure[$15 \bigotimes 7$]{
\includegraphics[width=0.485\textwidth]{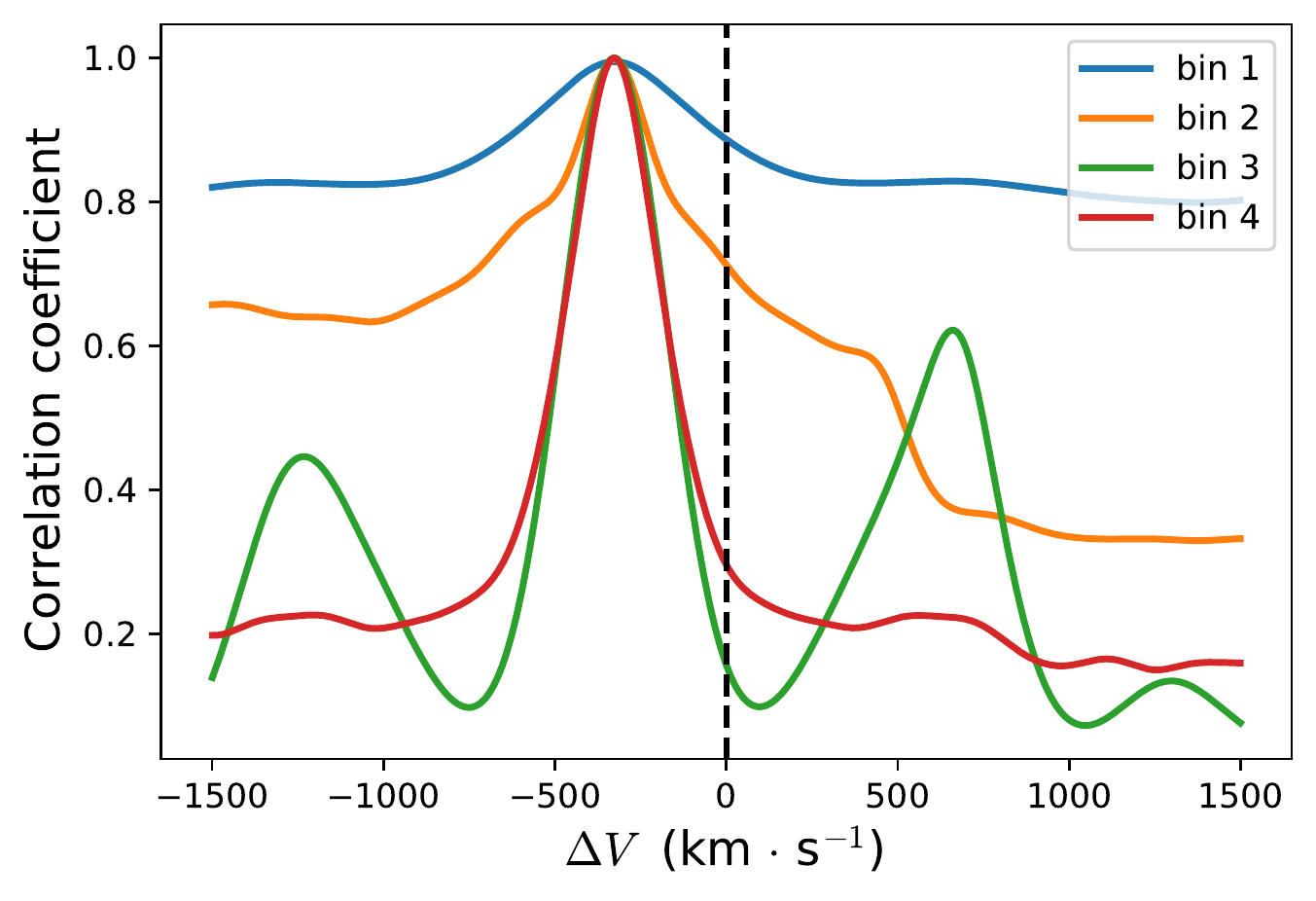}}
\caption{The correlation coefficient $r(\dv)$ plotted as a function of velocity difference $\dv$ between a pair of spaxels, $A=7$ and $B=15$ from the simulation based on  the central spaxel of MaNGA galaxy 7991-12701. The left and right panels show $r_{A^sB}$ and $r_{B^s A}$ respectively for $4$ bins, calculated using \eqref{eq:rab}. All the bins have maxima at similar $\dv$ in either panel and at the opposite sign value of the other panel -- important crosschecks. 
}
\label{fig:demo_cc}
\end{figure}

Figure~\ref{fig:demo_cc} displays the correlation coefficient as a function of the velocity difference, $\dv$, between the pair of spaxels. The left and right panels show $r_{A^sB}$ and $r_{B^s A}$ respectively for $4$ bins. 
For this illustration we use our baseline $\D=1.5$ \AA\ and $\nit=10$, later showing 
robustness to variation of these parameters. 
In both panels we find that $r$ exhibits global maxima at very similar values of $\dv$ (taking proper account of the sign), 
not only for $A^sB$ and $B^sA$, but also for 
different wavelength bins. Importantly, although the second and the fourth bin do not contain any strong feature, as evident from figure~\ref{fig:demo_spectra_cc}, 
the correlation $r$ shows a prominent peak at the correct $\dv$ values; this arises from the presence of many small features in these bins, and holds for both panels of 
figure~\ref{fig:demo_cc}. This illustrates the key benefit 
of this method in that it does not rely on single or few strong identifiable features in the spectra.

The maxima in the left and right panels arise at near-identical absolute values of $\dv$, i.e.\
$\dv_{BA} \approx - \dv_{AB}$ for all the bins as expected for an  accurate shift estimation. Since both the spectra are derived from the same spectra (the central one of MaNGA 7991-12701) and have little noise, all the maxima in this first example have very high values for the correlation coefficient, $r \approx 1$.
Following the algorithm described above in section \ref{sec:algorithm}, we find all 4 bins satisfy the selection criteria. Thus we have $4 \times 2=8$ `good' estimations of $\dv$ from both the panels. 

(Note that the secondary maxima in the third 
bin are readily identified as due to 
correlation between the $H\alpha$ line (6564.6~\AA\ 
restframe) and the NII line (6585.3~\AA\ restframe), 
giving a $\pm944$ km/s offset relative to the truth. This 
is an example of using not just statistics, 
but astrophysical knowledge, in assessing the 
data.)

From these eight maximum correlations we compute 
the mean and standard deviation for 
the velocity difference. The final result is $\dv_{AB} \approx 326.94 \pm 0.50$ \kmps\ (for our baseline $\D=1.5$ \AA, $\nit=10$, and 4 bins). This is wholly consistent with the simulation input $\dv_{7,15}= 326.83$ km/s, 
demonstrating validation for this first test.

In table~\ref{table:demo_sim_manga_7991-12701_spaxs}, we present the estimations of $\dv$ for the same pair of spaxels as in figure \ref{fig:demo_cc} but for various choices of $\D$, $\nit$, and $N_B$ (number of bins). All our estimations corresponding to these different choices are highly accurate with very small uncertainty  and consistent with each other demonstrating the robustness of this approach. We choose as our baseline $\D=1.5$ \AA, $\nit=10$, and $N_B=4$ for the best combination of accuracy, precision, and computational time efficiency.

\begin{table*}
\begin{tabular}{|c|c|c|c|c|}
 \hline
 \multicolumn{1}{|c|}{\multirow{3}{*}{$\Delta$ (in \AA)}} & \multicolumn{4}{|c|}{Estimated velocity difference ($\dv$) in \kmps}\\\cline{2-5}
& \multicolumn{2}{|c|}{$4$ bins}& \multicolumn{2}{|c|}{$8$ bins}\\\cline{2-5}
 & $\nit=10$ & $\nit=20$ &  $\nit=10$ & $\nit=20$\\
 \hline
 $1.5$  & $ 326.9 \pm 0.5 $ & $ 326.8 \pm 0.8 $ & $ 326.8 \pm 0.6 $ & $ 326.7 \pm 1.0$  \\
 \hline
 
  $2.0$  & $ 326.8 \pm 0.6  $ & $ 326.6 \pm 1.1$ & $ 326.7 \pm 0.7 $ & $326.5 \pm 1.2 $  \\
 \hline
 
  $3.0$  & $ 327.0 \pm 0.8 $ & $ 326.9 \pm 0.7 $ & $ 326.8 \pm 0.6 $ & $326.7 \pm 0.6  $  \\
 \hline
 
  $4.0$  & $ 327.4 \pm 1.7 $ & $ 327.4 \pm 1.7$ & $ 327.0 \pm 1.5 $ & $ 327.0 \pm 1.4 $  \\
 \hline

 \hline

 \hline
\end{tabular}
\caption{Velocity difference between the spaxels 7 and 15 (from the simulation test in section~\ref{sec:1emission}) for different values of smoothing scale $\Delta$, number of iterations $\nit$, and number of bins $N_B$ considered. The true velocity difference between these two spaxels is $326.8$ km/s. The estimations of $\dv$ for all the above choices of $\lbrace \Delta, \nit, N_B \rbrace$ match extremely well with the truth. However, we choose $\D=1.5$ \AA, $\nit=10$, and $N_B=4$ as our base line for the best combination of accuracy, precision, and computational time efficiency. }
\label{table:demo_sim_manga_7991-12701_spaxs}
\end{table*}

\begin{table}
\centering
 \begin{tabular}{||c|c|c|c||} 
 \hline
 Pair of spaxels (A,B) & Our estimation of $\Delta V_{AB}$ in km/s & True $\Delta V_{AB}$ in km/s & residual in km/s\\ [0.5ex] 
 \hline\hline
 $( 1, 2 )$ & $ 5.90 \pm 0.25 $ & $ 6.21 $ & \!\!\!\!$ -0.31 \pm 0.25 $ \\ 
 \hline

  $ (1,21 ) $ & $ 453.15 \pm 0.69 $  & $ 452.59 $ & $ 0.56 \pm 0.69$   \\ 
 \hline
 
 $( 2, 13 )$ & $ 330.27 \pm 0.78 $ & $ 329.91 $ & $ 0.36 \pm 0.78 $ \\
 \hline
 
 $( 4, 21 )$ & $ 432.08 \pm 0.70 $ & $ 431.20 $ & $ 0.88 \pm 0.70 $ \\
 \hline
 
  $( 5, 6 )$ & $ 12.79 \pm 0.70 $ & $ 13.18 $ & \!\!\!\!$ -0.39 \pm 0.70 $  \\
  \hline
 
  $ (7,11 ) $ & $ 163.76 \pm 0.44  $  & $ 163.48 $   & $ 0.28 \pm 0.44$  \\ 
 \hline
 
  $ ( 7,15) $ & $ 326.94 \pm 0.50  $  & $ 326.83 $   & $ 0.11 \pm 0.50$  \\ 
 \hline
 
 $( 9, 14 )$ & $ 251.14 \pm 0.78 $ & $ 251.37 $ & \!\!\!\!$ -0.23 \pm 0.78 $  \\
 \hline
 
 $( 10, 21 )$ & $ 292.45 \pm 0.34 $ & $ 291.98 $ & $ 0.46 \pm 0.34 $ \\
 \hline
  $ (11,15 ) $ & $ 163.01 \pm 0.85 $  & $ 163.34 $    & \!\!\!\!$ -0.33 \pm 0.85$ \\ 
 \hline
 
  $ (11,17 ) $ & $ 194.32 \pm 0.71 $  & $ 193.96 $    & $ 0.36 \pm 0.71$ \\ 
 \hline
 
   $ (13,19 ) $ & $ 104.20 \pm 0.89 $  & $ 103.51 $    & $ 0.69 \pm 0.89$ \\ 
 \hline
 
 $( 16, 19 )$ & $ 31.15 \pm 0.31 $ & $ 31.55 $ & \!\!\!\!$ -0.40 \pm 0.31 $ \\
  \hline

 \hline

 \hline
\end{tabular}
\caption{We compare our estimation of velocity difference $\Delta V_{AB}$ between various pair of spaxels with the corresponding truth values used in the simulation. Note that spaxel 11 is the central spaxel in the simulation, so (1,21) spans from one edge of the galaxy to the other. 
}
\label{tab:demo_sim_set7_spaxs}
\end{table}

To assess the technique further we estimate the velocity difference 
for several other pairs of spaxels spanning the range of 21 spaxels (with spaxel 11 the central one). 
In table~\ref{tab:demo_sim_set7_spaxs}, our estimated $\dv$ values are compared with the corresponding true velocity differences used in the simulation. Again we find excellent matches between our estimations and the truths, with uncertainty below 
$1$~\kmps~ in all these cases. Note that such a small uncertainty arises due to the homogeneity of the spectral properties and (low) noise across all the spectra. 
The remaining subsections of this section will validate the method on simulations under more challenging observational circumstances, before proceeding to real observed data in section~\ref{sec:data}.

\subsubsection{Combining information from different spaxel pairs}
\label{sec:combine} 

Before proceeding further  we briefly explain here 
how we construct the galaxy rotation curves. 
Without loss of generality, we can assume the 
central spaxel has zero rotational velocity, i.e.\ 
we compute the velocity of each spaxel on 
the diameter (more precisely along 
the major axis of the galaxy) relative to the central velocity and  construct the galaxy rotation curve \footnote{ By setting the velocity of the central spaxel to zero, without loss generality, we are essentially measuring the LoS velocities of the spaxels relative to the reference central spaxel. In reality, the central spaxel may not be identified. In such cases one can assume the symmetry in the rotation curve to determine the centre and estimate the regularised spaxel velocities.
}. 

However, the spectrum of the central spaxel itself could be noisy 
or distorted, so to avoid such an issue we use the estimated velocity differences between {\it all\/} the pairs, 
i.e.\ $\dv_{i,j}$ where $i,j$ represents all the spaxels (at least all 
the spaxels with good 
S/N)\footnote{When 
spaxels have poor S/N we still include them as 
$i$ spaxels but not as $j$ spaxels, i.e.\ we 
do not measure good spaxels relative to bad. 
This does not occur for tests in 
section~\ref{sec:validsim} 
but see section~\ref{sec:grcmanga} for further 
details.}. 
We then fit for the velocities of all the spaxels 
simultaneously using all the $\dv_{i,j}$ measurements, i.e.\ all spaxel pairs. 
Defining the velocity of the central spaxel to be zero, 
we thus have $N_{\rm spax}-1$ free parameters in the fit, and up to 
$N_{\rm spax}(N_{\rm spax}-1)/2$ measurements, where  $N_{\rm spax}$ is the total number of spaxels. 
Since $N_{\rm spax}$ can be large in principle (e.g.\ around a few thousand for the MaNGA IFU observations on a 2D grid), we employ  Hamiltonian Monte Carlo (HMC) sampling with the help of the pystan package \citep{stan_2017,pystan}. 
The fit takes care of the consistency between the velocities of the spaxels, 
i.e.\ the triangle inequality 
$\vec\dv_{AB}+\vec\dv_{BC}=\vec\dv_{AC}$, 
or alerts us to inconsistency by not converging or giving a poor $\chi^2$.

\subsubsection{Constructing the galaxy rotation curves} 
\label{sec:1emission_grc} 

Now we complete the construction of the galaxy rotation curve for the galaxy in the case where the spectra in all 21 spaxels are generated from the observed spectrum in the central spaxel of MaNGA 7991-12701. 
All pairs of spaxels have good 
signal to noise ratio (S/N $\sim 60$ as evident from figure \ref{fig:S_N_lowNoise}) so we include the $\dv$ estimated for all the pairs into the fit. 

\begin{figure}
\centering
\subfigure[Emission line dominated (section \ref{sec:1emission})]{
\includegraphics[width=0.485\textwidth]{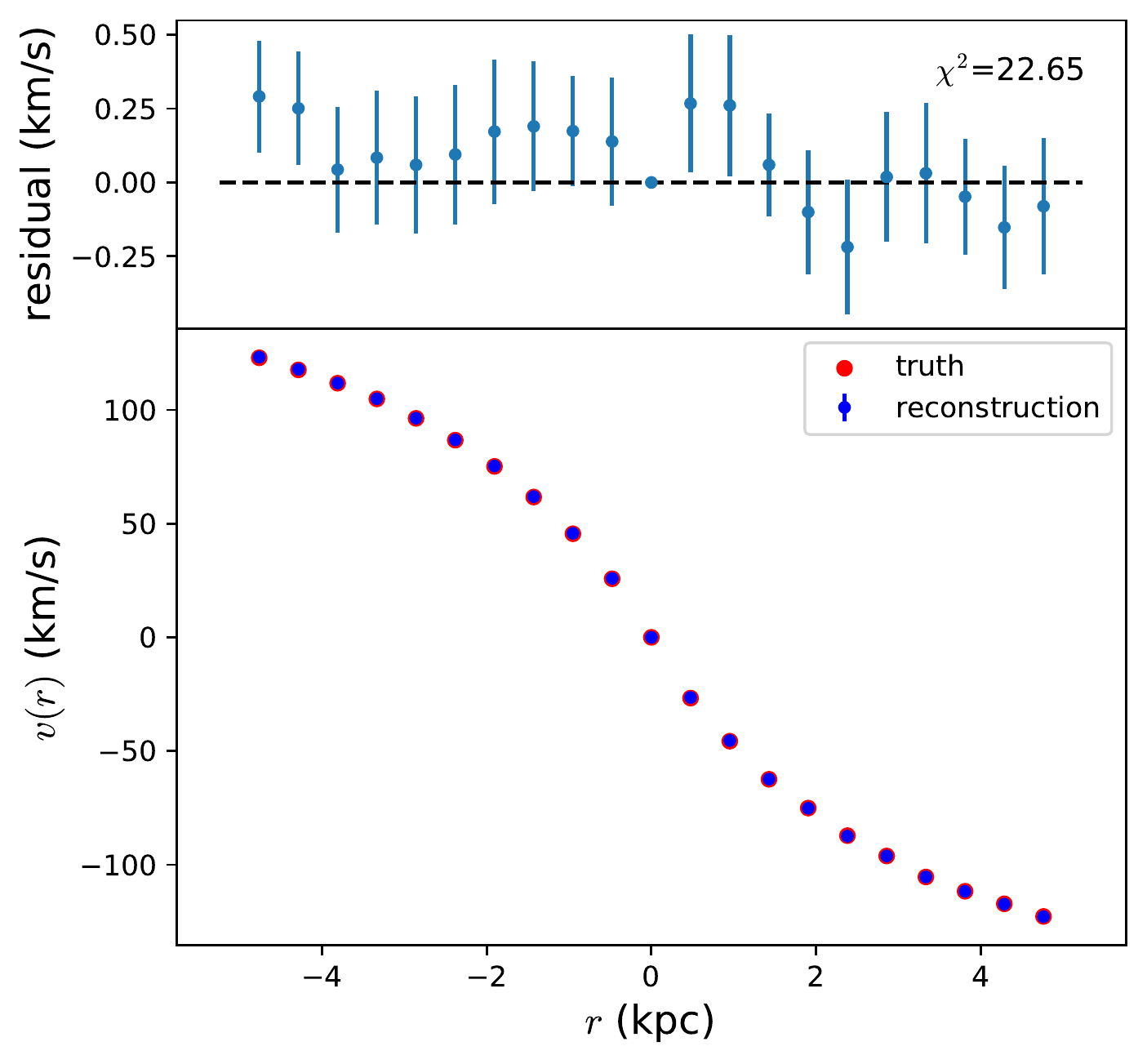}}
\subfigure[Absorption line dominated (section \ref{sec:1absorption})]{
\includegraphics[width=0.485\textwidth]{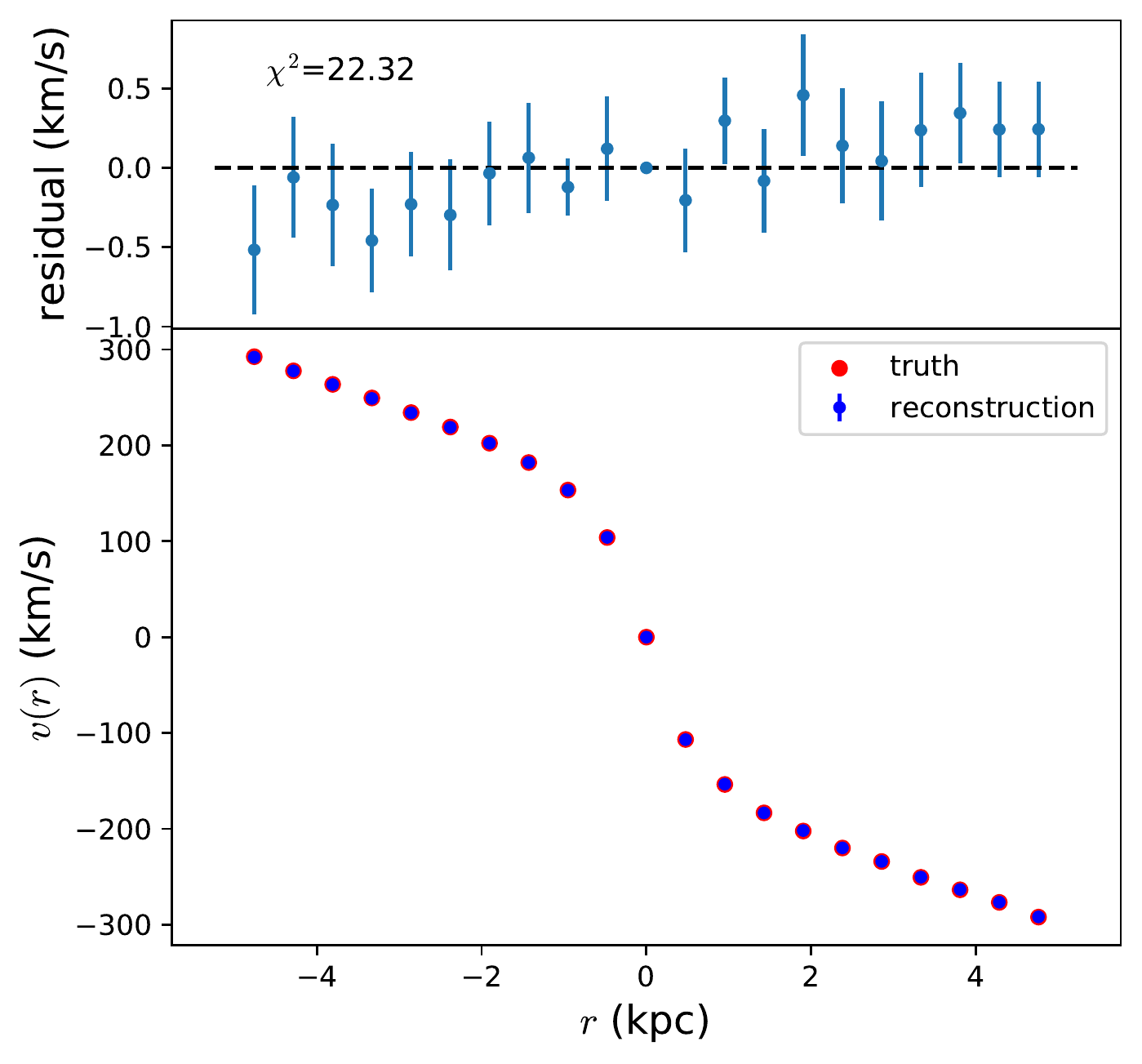}}
\subfigure[Inhomogeneous spectral properties:\protect\newline  \hspace*{1.8em} emission line dominated  (section \ref{sec:2emission})]{
\includegraphics[width=0.485\textwidth]{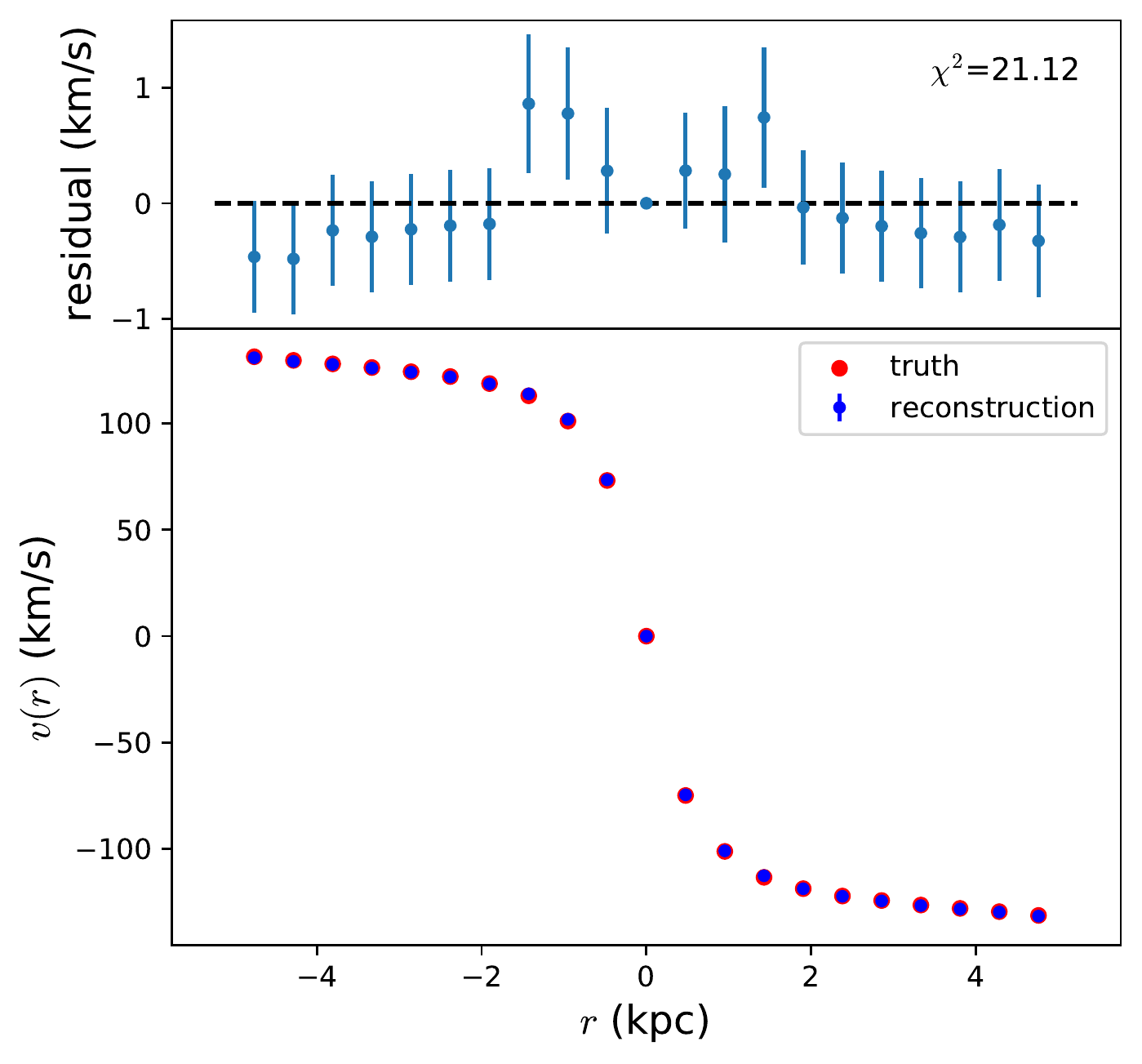}}
\subfigure[Inhomogeneous spectral properties:\protect\newline  \hspace*{1.8em} weak emission vs absorption line dominated (section \ref{sec:absorption_vs_weak_emission})]{
\includegraphics[width=0.485\textwidth]{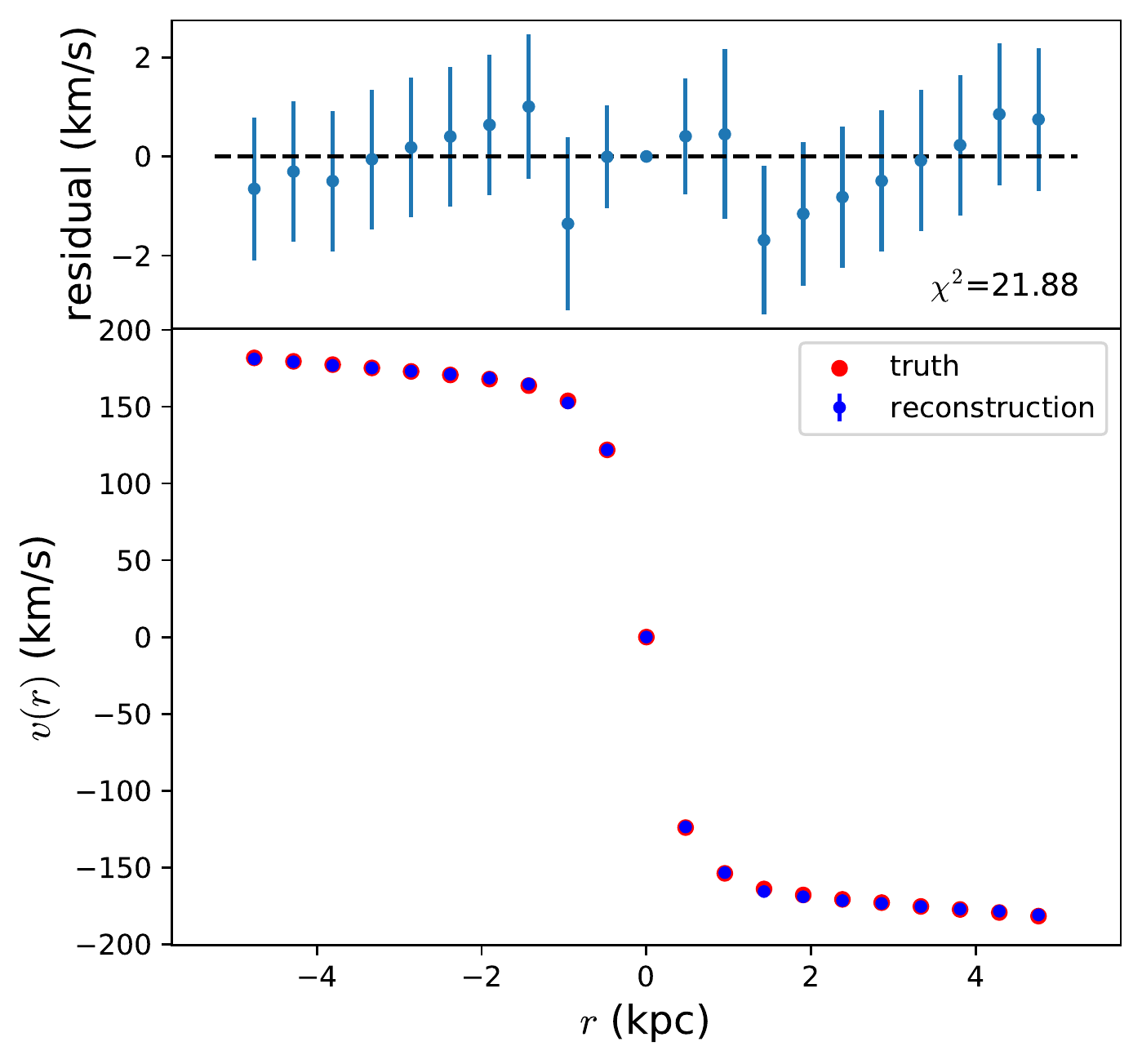}}
\caption{The four panels present the results of the tests described in sections \ref{sec:1emission} -- \ref{sec:absorption_vs_weak_emission} clockwise from the top-left. The top two panels compare the our estimated galaxy rotation curves with the corresponding truths when the spectra in different spaxels are generated from a single base spectrum -- top left panel {\bf (a)}: emission line dominated (the orange spectrum in 
figure~\ref{fig:emission_spectra_7991-12701}), top right panel {\bf (b)}: absorption line dominated (the blue spectrum in figure~\ref{fig:absorption_spectra_8952-9102}). The bottom two panels corresponds to the tests when the spectral properties change like a step function along the major axis as described in sections \ref{sec:2emission} and \ref{sec:absorption_vs_weak_emission}. 
The analyses are carried out blindly and when the truths are revealed we find that the velocities are recovered accurately for all the cases. The uncertainty in $v(r)$ estimation is so small that the errorbars are not visible in the plots. The subplot at the top of each panel shows the residual for each estimation along with the errorbars. The residuals are very small, especially for the top two cases with homogeneous spectral properties. The $\chi^2$ values for all four cases are close to $20$, as expected for $20$ degrees of freedom. Note that there is 
(modest) correlation between the fit velocities; the $\chi^2$ estimation considers the whole covariance matrix, not just the diagonal terms.}
\label{fig:blind_tests}
\end{figure}

 To ensure the method is fairly tested, we conduct each test in pairs considering two realizations 
(as described above in equation~\ref{eq:vrmodel}),
each with independent input parameters and noise realizations. The first test is conducted such that the input rotation curve is known, and then followed up by a second test which is similar except the input rotation curve is not revealed until the results have been collected (a blind test). 
In the top left panel of figure \ref{fig:blind_tests} we compare our estimated galaxy rotation curve with the true one for the blind case\footnote{The results for the `known cases' are statistically same as the blind cases and, therefore, are not shown separately in this article.}.
We find an excellent match between our estimations and the true velocities. The uncertainties in estimations are so small 
($\sim0.1$ km/s) 
that they are not visible in the $v(r)$ plot. The subplot at the top of the panel shows the residuals of our estimations, which are typically $\mathcal{O}(0.1)$ \kmps~. Since all the $\dv_{ij}$ estimations, which are correlated, are considered in the fit, 
``chi-by-eye'' is not accurate. 
We compute 
$\chi^2 \equiv \sum_{ij}({\bf V_{\rm est}-V_{\rm true}})_i \cdot {\bf Cov}^{-1}_{ij} \cdot ({\bf V_{\rm est}-V_{\rm true}})^T_j$ where ${\bf Cov}$ is the covariance matrix of the fit velocities. One expects $\chi^2 \sim 20$ since the velocity of the central spaxel out of $21$ is fixed to zero. 
The $\chi^2$ value is 
$22.29$, indicating  
that our method is both 
accurate and statistically robust for 
estimating the galaxy rotation curve. 

In the following sections we increase the difficulty 
by changing the spectrum, noise, and adding inhomogeneity.


\subsection{Absorption line dominated spectra} \label{sec:1absorption} 

Next we test how our method performs when the galaxy spectra do not contain any sharp emission line (in other words, if the galaxy is not star-forming), rather they are 
dominated by absorption lines. We again simulate spectra in $21$ spaxels along the major axis of a galaxy. For simplicity, the spectral properties for all spaxels here are fixed to that of the central spaxel $(32,32)$ of the MaNGA galaxy 8592--9102, shown in 
figure~\ref{fig:absorption_spectra_8952-9102} by the blue colour. The comparison between our estimation of rotational velocities of the spaxels with the truth is presented in the top right panel of figure~\ref{fig:blind_tests}. Again we find an excellent match with the truth and the small errorbars are not visible in the $v(r)$ plot. The residuals, shown in the top subplot, are again quite small, 
however the residuals and their  uncertainties here are slightly larger as compared to the case with the emission line dominated spectra because the 
emission lines are typically stronger than the absorption lines. E.g., the average residual in the absorption line dominated case is $\sim 0.22$ \kmps~ which is slightly larger compared to that of the emission line dominated case ($\sim 0.14$ \kmps). The $\chi^2$ value for this case (taking the full covariance matrix of the fit velocities into account) $22.32$ for 20 degrees of freedom, is again reasonable.

\subsection{Different spectral properties: emission line dominated }
\label{sec:2emission}

In the previous two subsections, we tested the method on the simple scenarios when the spectral properties are kept the same across all the spaxels 
(only shifted according to the velocity differences along line of sight). But in reality, galaxies can go from having absorption to emission lines as one moves across them, because of stellar population gradients.
So here we test the scenarios when the spectral properties 
are inhomogeneous. To challenge the technique, we take a radical step function in radius, where ``inner'' 
spaxels have a different spectrum than ``outer'' spaxels. 
This and the next subsection take two different models for the base spectra.

Here, both the inner and outer spaxels are simulated from  
strong emission line dominated spectra, 
but now only the inner spaxels come from the central spaxel ($37,37$) 
of the MaNGA galaxy 7991--12701, 
while the outer spaxels are simulated 
with the spaxel ($37,12$) of 7991--12701.  
These two emission line dominated model spectra have been shown in figure~\ref{fig:emission_spectra_7991-12701}. While they have the same emission lines, their continua are quite different; the former has red slope continuum whereas the latter has a bluer continuum. This subsection tests our method on this case of spectral property inhomogeneity.

We compare our estimation of rotational velocities (of the spaxels) against the true velocities for this scenario in bottom left panel of figure \ref{fig:blind_tests}.
Since there is a sharp transition of the spectrum, we get slightly stronger correlations in the fit velocities among the inner spaxels (8-14, the central one is spaxel 11) and the outer ones. This leads to the pattern apparent
in the residual plot shown at the top subplot. Nevertheless, even in this extreme step function case 
the $\chi^2$ is $21.12$ for 
$20$ degrees of freedom.

\subsection{Different spectral properties: absorption line vs weak emission line dominated spectra}
\label{sec:absorption_vs_weak_emission}

Next we test the method with simulations again involving a step function in spectral properties, 
now with the inner spaxels based on weak absorption line dominated spectra and the outer spaxels based on weak emission line dominated spectra. 
Specifically, we use the (32,37) spaxel 
of the MaNGA galaxy 8952--9102, shown as the orange spectrum (with weak absorption lines) in 
figure~\ref{fig:absorption_spectra_8952-9102}, and  the (32,43) 
spaxel of the same galaxy,  the green 
spectrum (with weak emission lines) in that figure. 
Note that these two base spectra have much weaker features compared to the emission line dominated spectra shown in 
figure~\ref{fig:emission_spectra_7991-12701} and analysed in 
section~\ref{sec:2emission}. 

The comparison between our estimation of the rotational velocities 
with the true ones appears in the bottom right panel of 
\ref{fig:blind_tests}. 
Our estimation matches very well with the truth even with the step function 
variation in the spectral properties among spaxels where the features are weak and of 
different types (absorption vs emission). Again, due to the sharp transition in the spectra we get some structure in the fit as evident from the pattern in the residuals shown in the top subplot. 
Nevertheless, the $\chi^2$ value
is good: $21.9$. 
Thus, even in this more inhomogeneous case, the method succeeds.

\section{Validating on simulations with different noise levels}\label{sec:SN_test} 
\begin{figure}
\centering
\includegraphics[width=\textwidth]{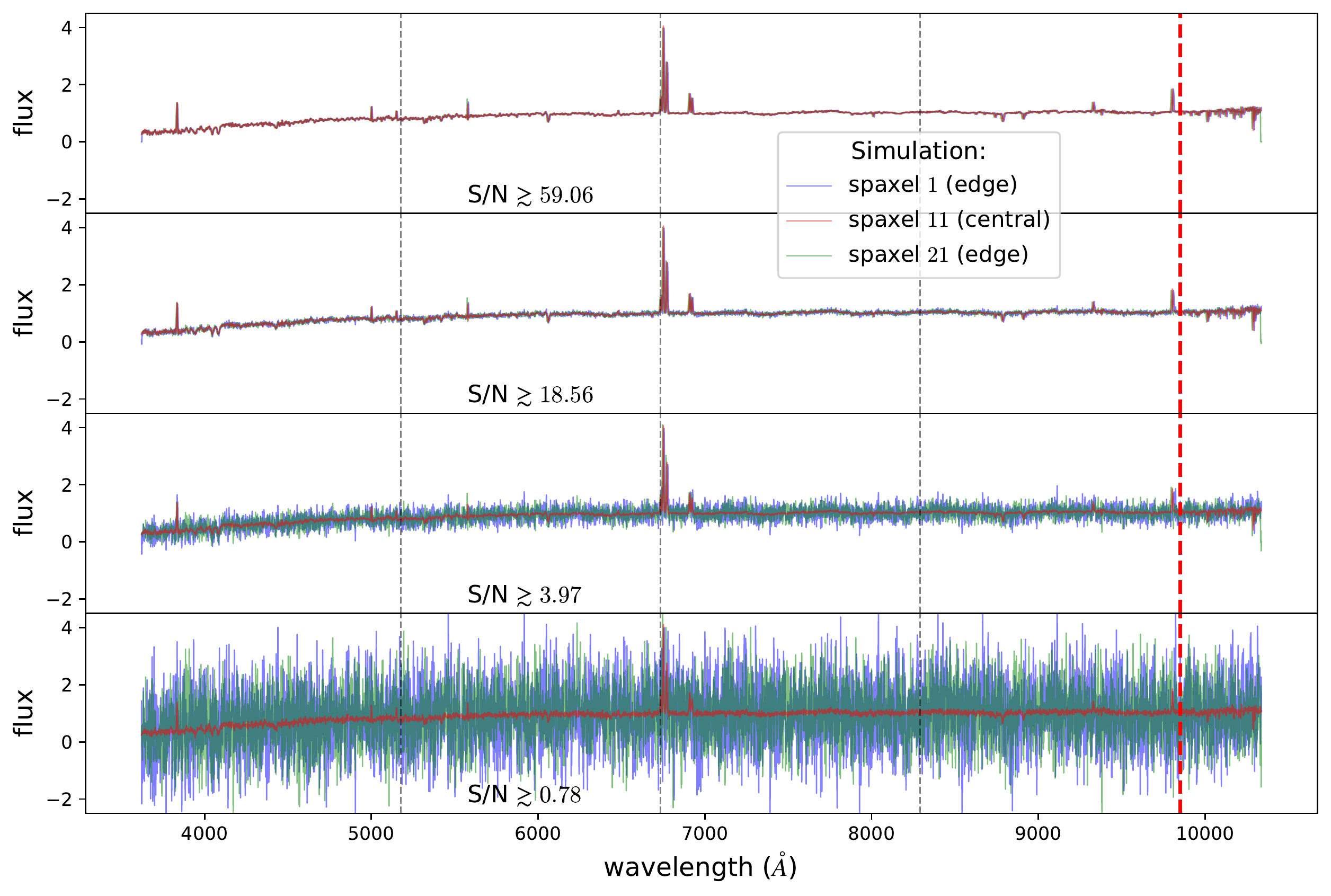}
\caption{Emission line dominated spectra are shown for four sets (used in section~\ref{sec:SN_test_emission}) with different noise levels -- low, medium, high, and very high -- from the top to the bottom panel.  Each panel shows spectra in three spaxels, two edge spaxels and the central one. S/N for these four sets are as low as $59.06,~18.56,~3.97,~0.78$, as shown in the respective panels.
}
\label{fig:SN_emission_spectra}
\end{figure}

\begin{figure}
\centering
\subfigure[Emission line dominated spectra]{
\includegraphics[width=0.485\textwidth]{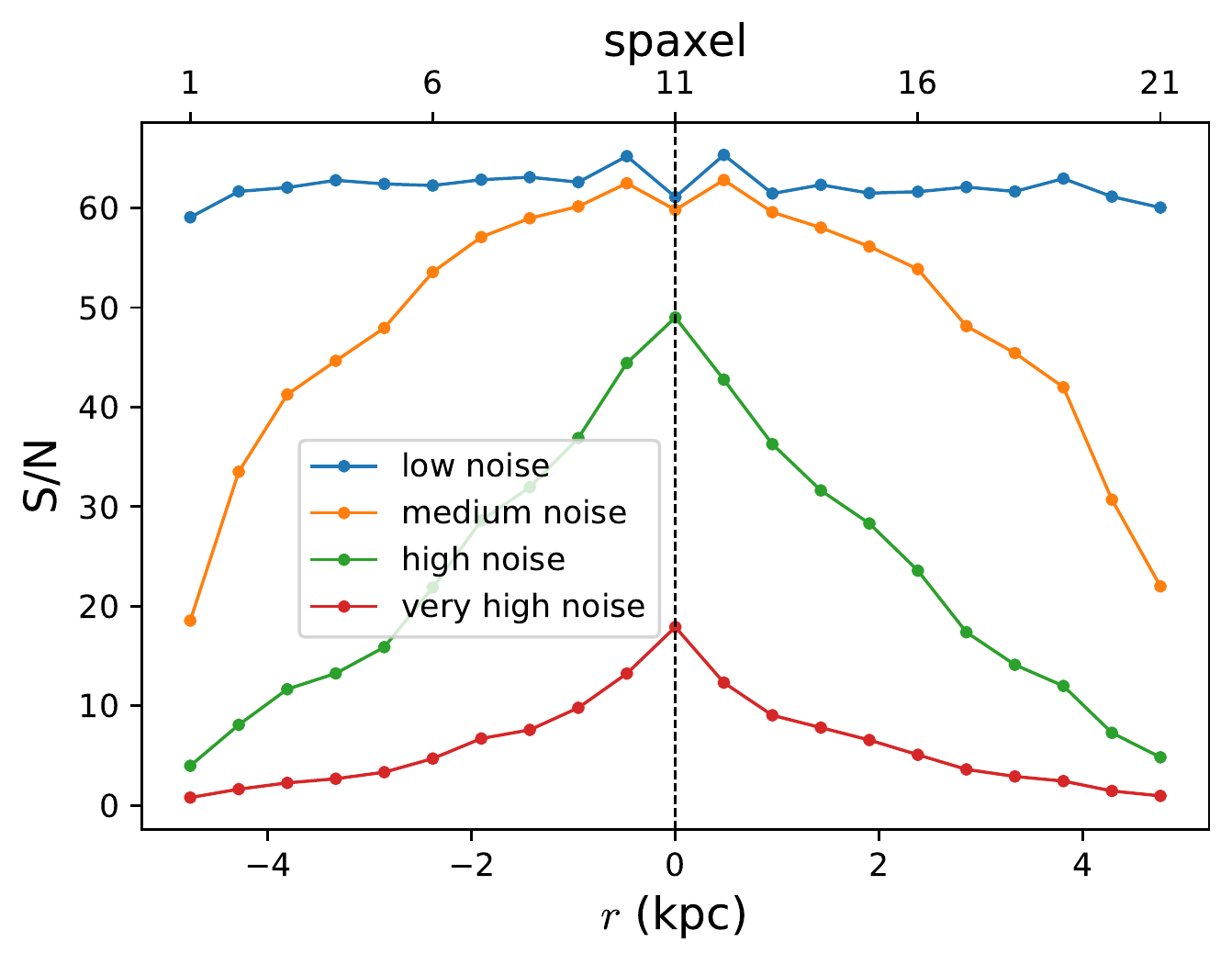}}
\subfigure[Absorption line dominated spectra]{
\includegraphics[width=0.485\textwidth]{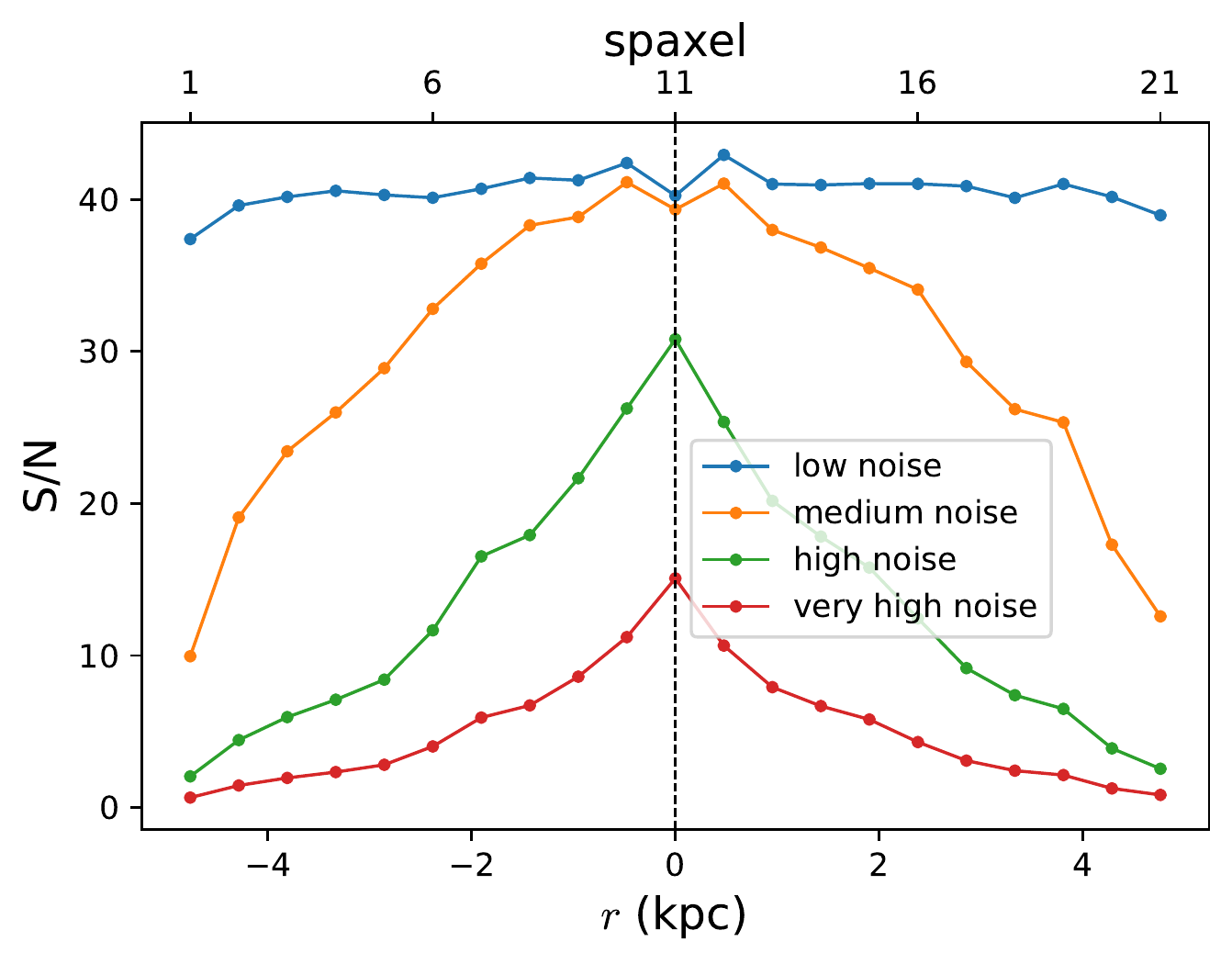}}
\caption{The variation of signal to noise ratio (S/N) of the spectra across the spaxels is shown for the four sets with different noise levels -- low, medium, high, and very high. 
The left panel is for 
simulations from an emission line dominated spectrum (section~\ref{sec:SN_test_emission}); the right panel for an absorption line dominated spectrum 
(section~\ref{sec:SN_test_absp}). Note that S/N of the three spaxels in each panel of figure \protect\ref{fig:SN_emission_spectra} are shown by the middle (central spaxel) and the two boundary dots (edge spaxels) on the corresponding curves in the left panel. Same applies for figure \protect\ref{fig:SN_absorption_spectra} and the right panel.
}
\label{fig:S_N_SNtest}
\end{figure}


In the previous section we apply the method to different scenarios while keeping the noise level low, giving high S/N in the spectra across the spaxels (S/N $\gtrsim 10$ in the worst cases). Because of that our estimations are not only in extremely good agreement with the truth in every case but have quite small uncertainty. In this section we systematically test the method on the spectra with different noise levels.

\subsection{Emission line dominated}  \label{sec:SN_test_emission} 

We again simulate spectra in the 21 spaxels generated from the same emission line dominated spectrum, namely of the central spaxel ($37,37$) of the MaNGA galaxy 7991--12701 (shown by orange in 
figure~\ref{fig:emission_spectra_7991-12701}), 
now with four different noise levels. 
Figure~\ref{fig:SN_emission_spectra} shows the spectra in three spaxels (two edge spaxels and the central one) for the four sets with four noise levels -- low 
(similar to in the previous section), medium, high, and very high -- from top to bottom panel. Notice that the spectra of the outer spaxels have much larger noise, because of being faint, most evident in the high and very high noise sets shown in the bottom two panels. The spectra in the edge spaxels still contain the sharp emission lines, however, the continua appear extremely noisy. The S/N in the spectra of the spaxels are shown in the left panel of 
figure~\ref{fig:S_N_SNtest}; note the rapid drop 
for the medium and especially for the high and very high noise sets 
going radially outwards. The spectra in the edge spaxels have S/N $\sim 4$ for the high noise set and S/N as low as $\sim 0.8$ for the very high noise set. Note that for all the sets we keep the true velocities the same so that the results can be compared easily.

\begin{figure}
\centering
\subfigure[low noise set (S/N $\gtrsim 59.06$)]{
\includegraphics[width=0.49\textwidth]{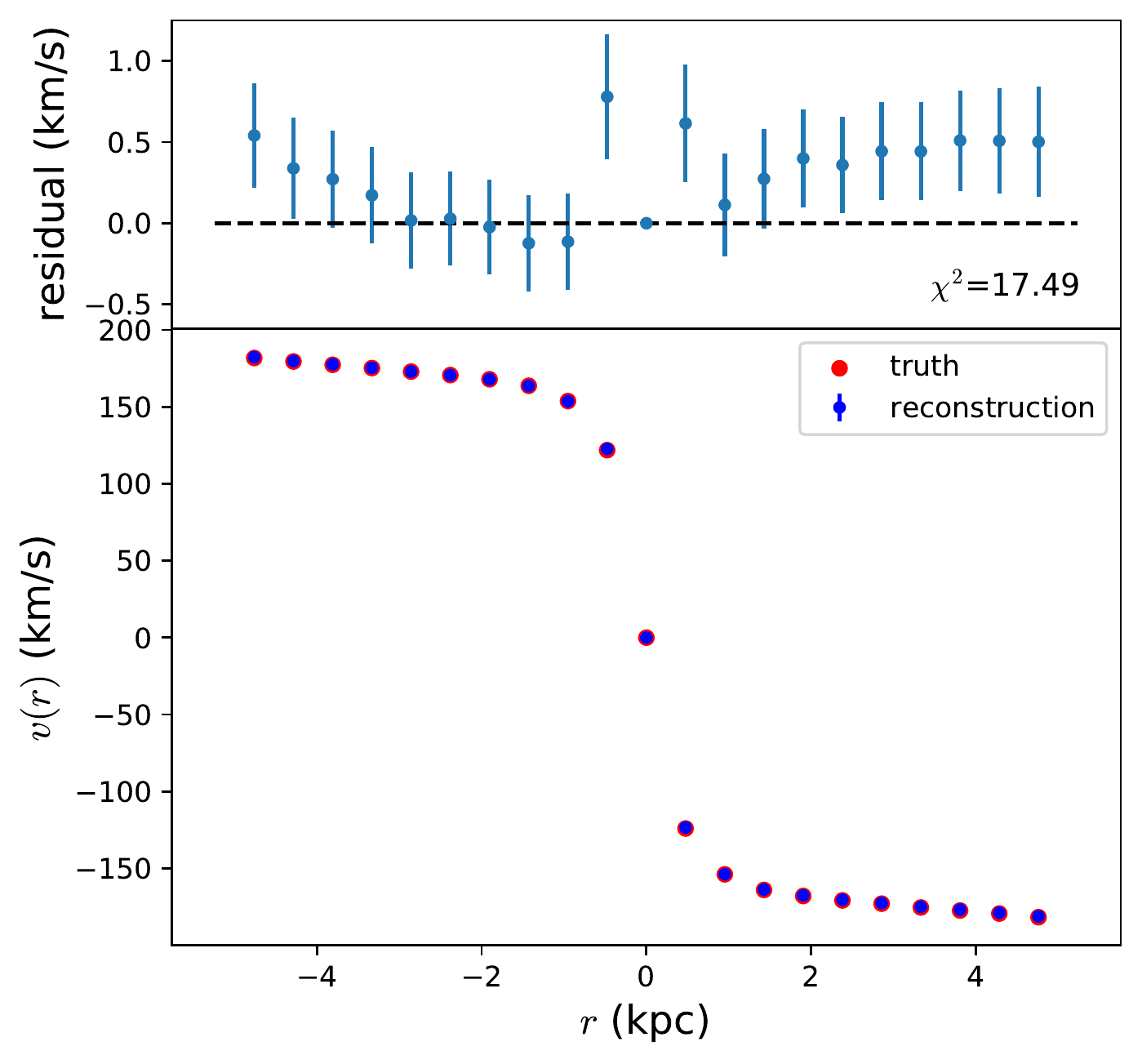}\hspace*{-2mm}}
\subfigure[medium noise set (S/N $\gtrsim 18.56$)]{
\includegraphics[width=0.49\textwidth]{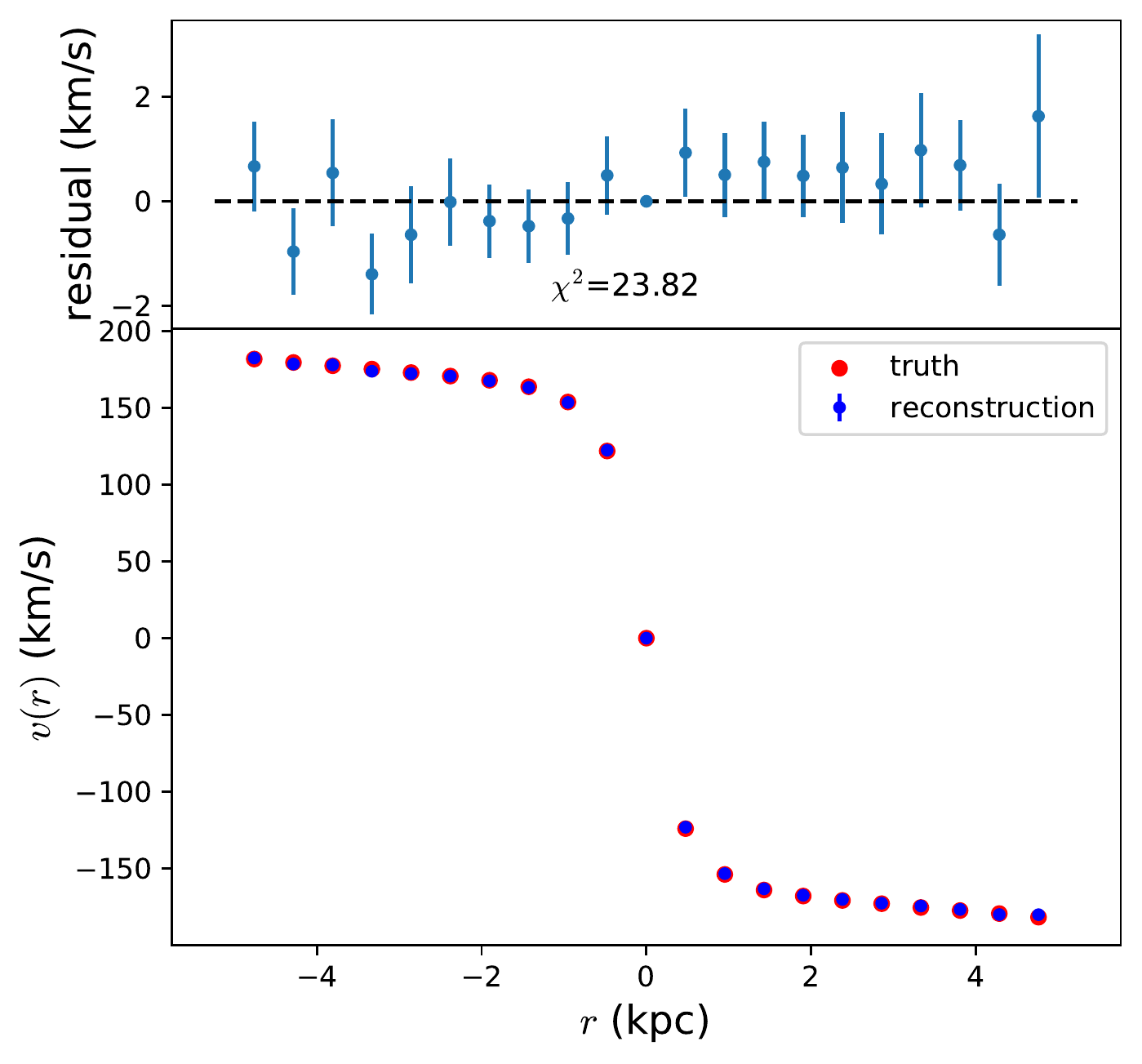}}\\
\subfigure[high noise set (S/N $\gtrsim 3.97$)]{
\includegraphics[width=0.49\textwidth]{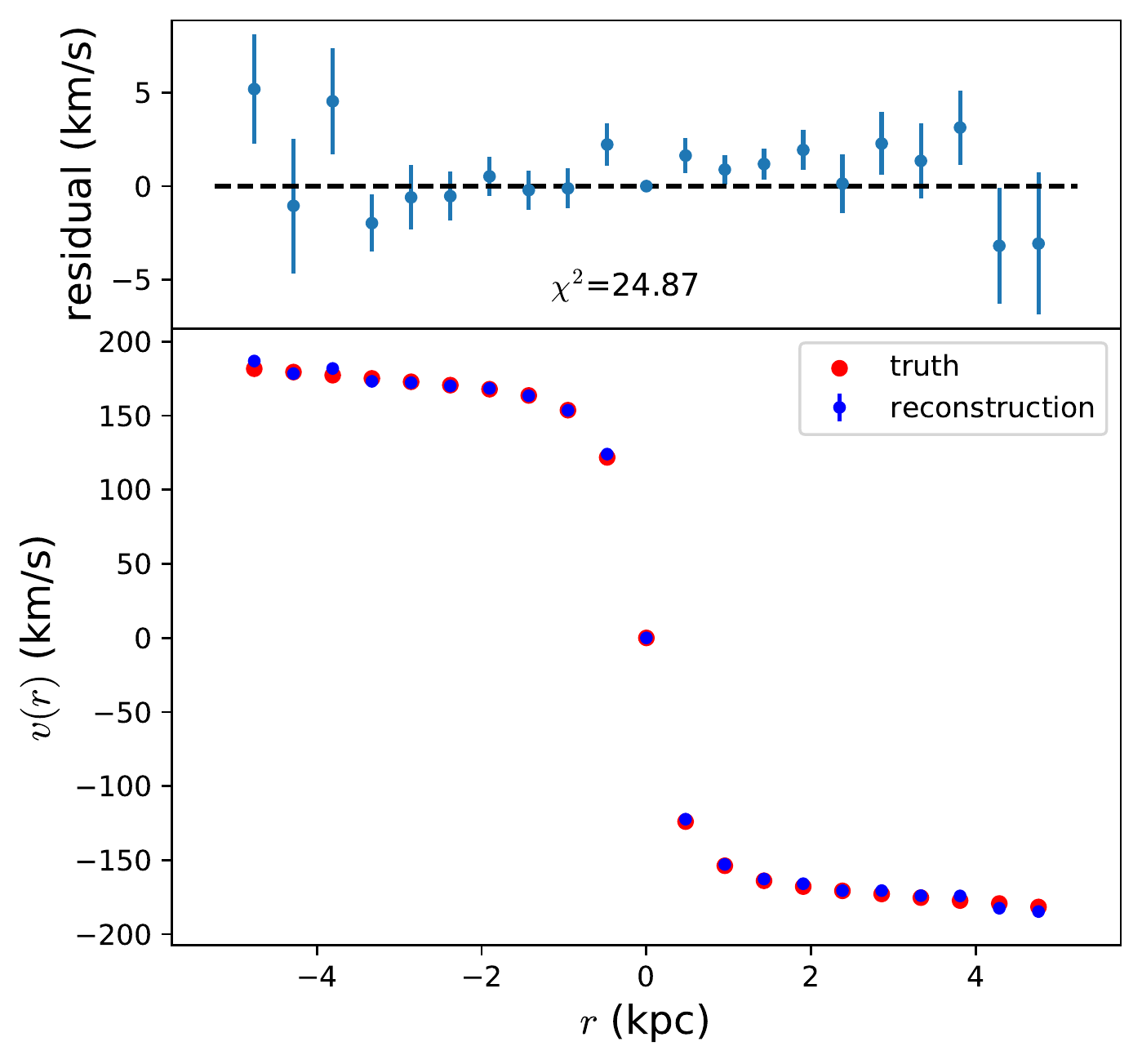}}\hspace*{-2mm}
\subfigure[very high noise set (S/N $\gtrsim 0.78$)]{
\includegraphics[width=0.49\textwidth]{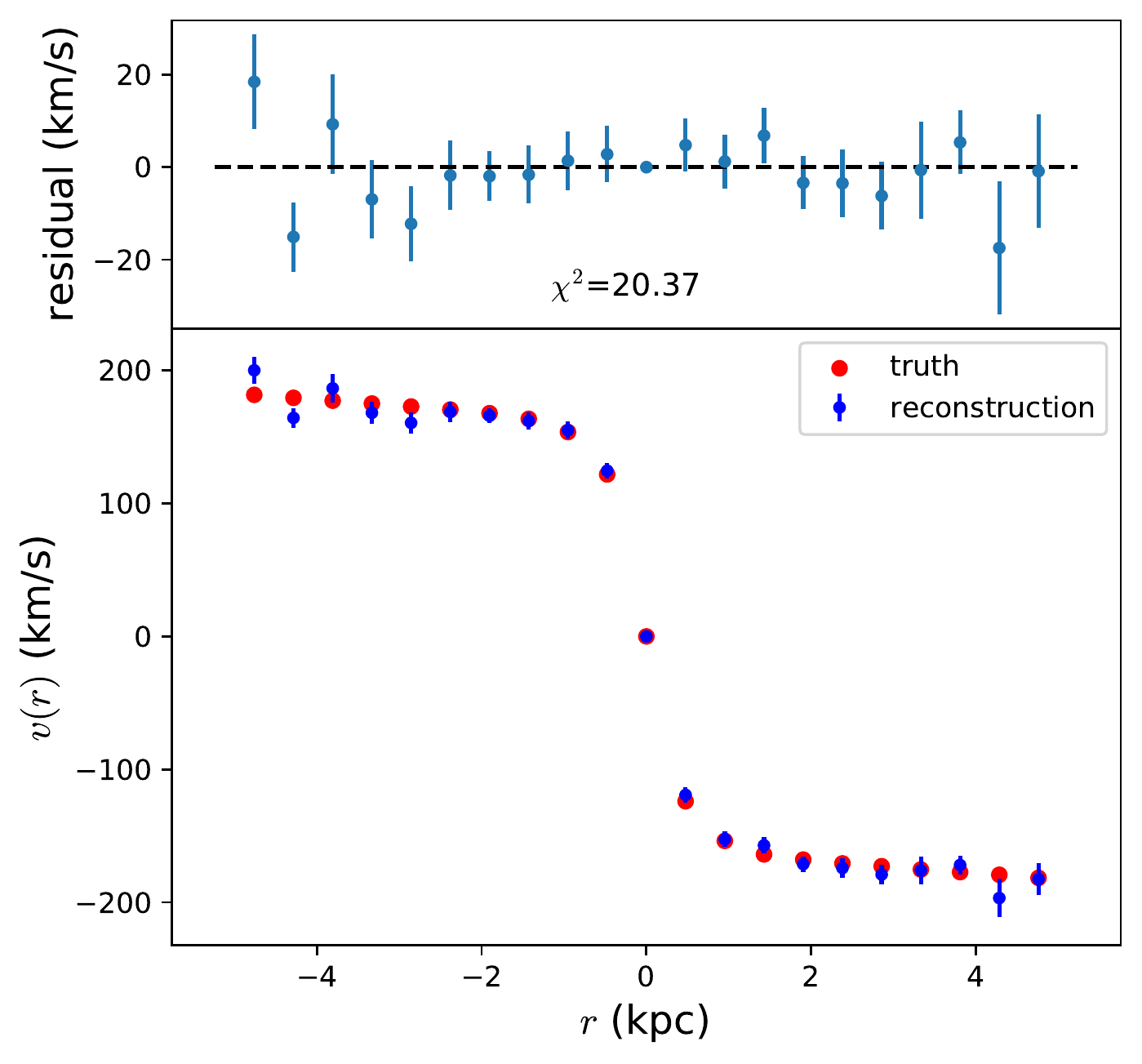}\label{fig:SN_test_emission_VHN}}
\caption{Velocity reconstruction for spaxels with  emission line dominated spectra, for  
different noise levels. 
}
\label{fig:SN_test_emission}
\end{figure}

For the higher noise sets we find that a slightly larger smoothing scale gives better accuracy and robust results. Thus, we use $\Delta=2.0$ \AA\ for the medium and higher noise sets while keep $\Delta=1.5$ \AA\ for the low noise set, same as before.
In the three panels of figure~\ref{fig:SN_test_emission} we compare the estimated rotational velocities for each of the four noise level sets with the true velocities (shown by the red dots). 
We find that our estimated  velocities for the low and medium noise levels virtually coincide with the truth, 
again with very little uncertainty. Remarkably, even for the higher noise level cases the velocities estimated are 
good, $\chi^2=24.9$ and $\chi^2=20.4$ -- mostly 
the outermost spaxels are slightly off 
and the uncertainties 
increase (recall that S/N drops down to $\sim 4$ or $\sim 1$ in the edge spaxels). 
Therefore, our method can estimate the rotational velocities accurately for the emission line dominated spectra even when the noise in the data is significant.

\subsection{Absorption line dominated}
\label{sec:SN_test_absp}  

\begin{figure}
\centering
\includegraphics[width=\textwidth]{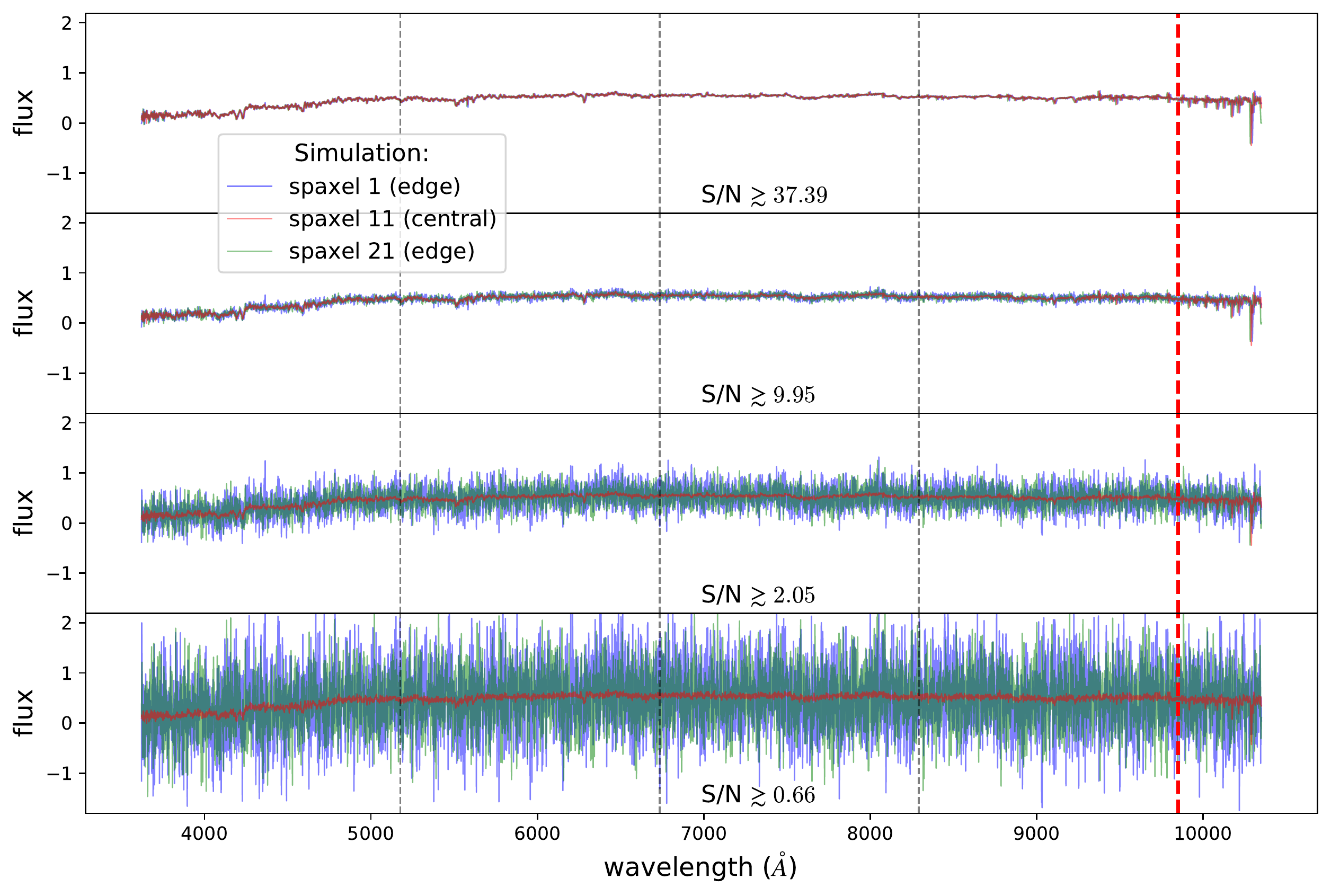}
\caption{Absorption line dominated spectra are shown for four sets (used in section~\ref{sec:SN_test_absp}) with different noise levels -- low, medium, high, and very high -- from the top to the bottom panel. Each panel shows spectra in three spaxels, two edge spaxels and the central one. S/N for these four sets are as low as $37.39,~9.95,~2.05,~0.66$, as shown in the respective panels. Notice that in the bottom two panels barely any feature is visible.
}
\label{fig:SN_absorption_spectra}
\end{figure}

Next we test the method on absorption line dominated spectral data with four sets having different noise levels. 
We use the weak absorption line dominated model spectrum from the $(32,37)$ spaxel of the MaNGA galaxy 8952--9102, shown by orange colour in 
figure~\ref{fig:absorption_spectra_8952-9102}. The spectra in three spaxels (the two edge spaxels and the central one) are shown in 
figure~\ref{fig:SN_absorption_spectra}, for the low, medium, high, and very high noise level cases. 
The right panel of figure~\ref{fig:S_N_SNtest} portrays how the S/N of the spectra varies across the spaxels for the four sets; again, S/N decays significantly at the outer-disk spaxels for the noisier sets, reaching 
as low as 
$\sim 2$ and $\sim0.7$ in the two highest noise cases. 
Indeed, it is difficult to discern 
by eye any clear features 
in the bottom panels of figure~\ref{fig:SN_absorption_spectra}. 

\begin{figure}
\centering
\subfigure[low noise set (S/N $\gtrsim 37.39$)]{
\includegraphics[width=0.49\textwidth]{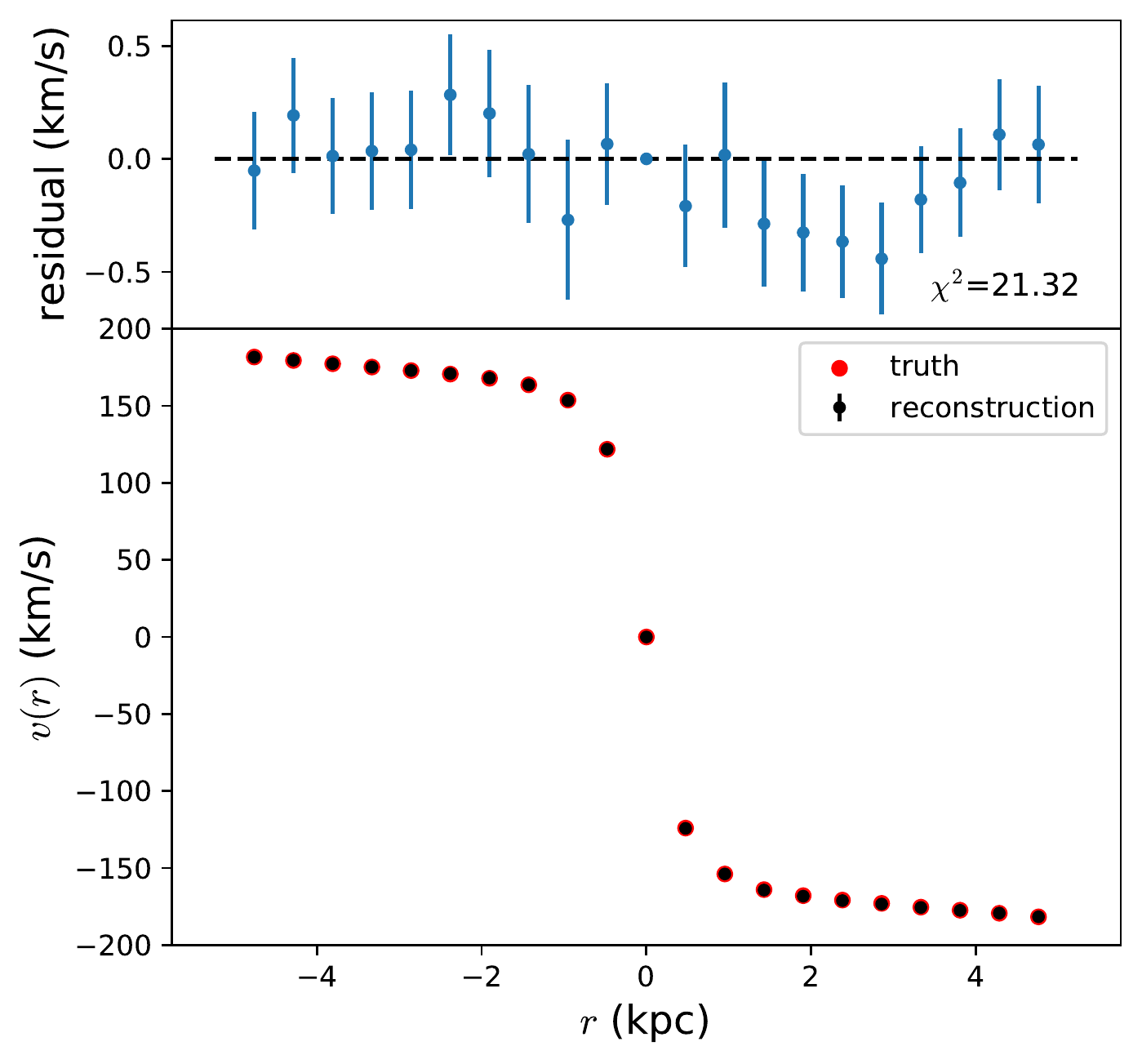}\hspace*{-2mm}}
\subfigure[medium noise set (S/N $\gtrsim 9.95$)]{
\includegraphics[width=0.49\textwidth]{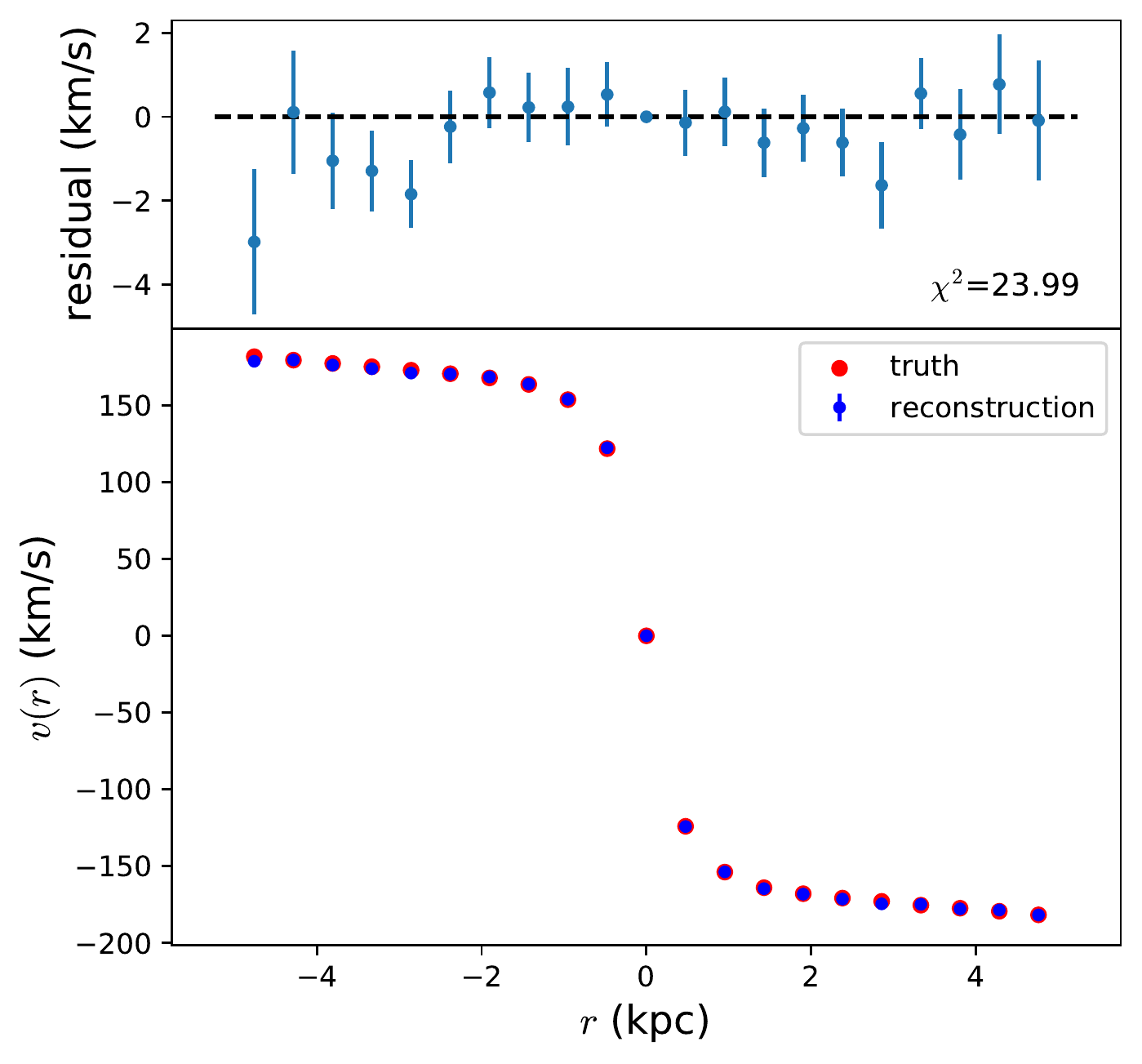}}\\
\subfigure[high noise set (S/N $\gtrsim 2.05$)]{
\includegraphics[width=0.49\textwidth]{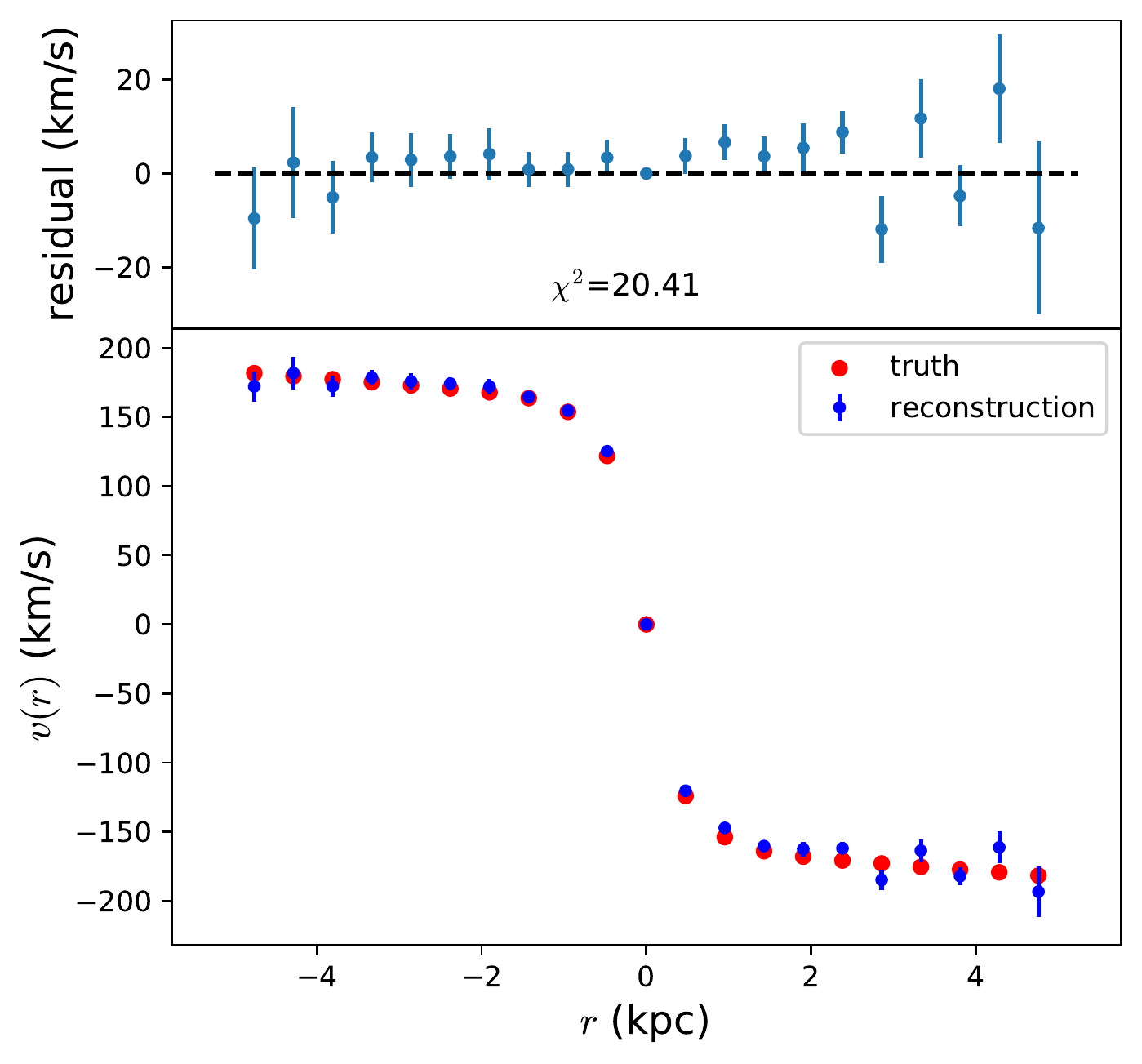}\hspace*{-2mm}}
\subfigure[very high noise set (S/N $\gtrsim 0.66$)]{
\includegraphics[width=0.49\textwidth]{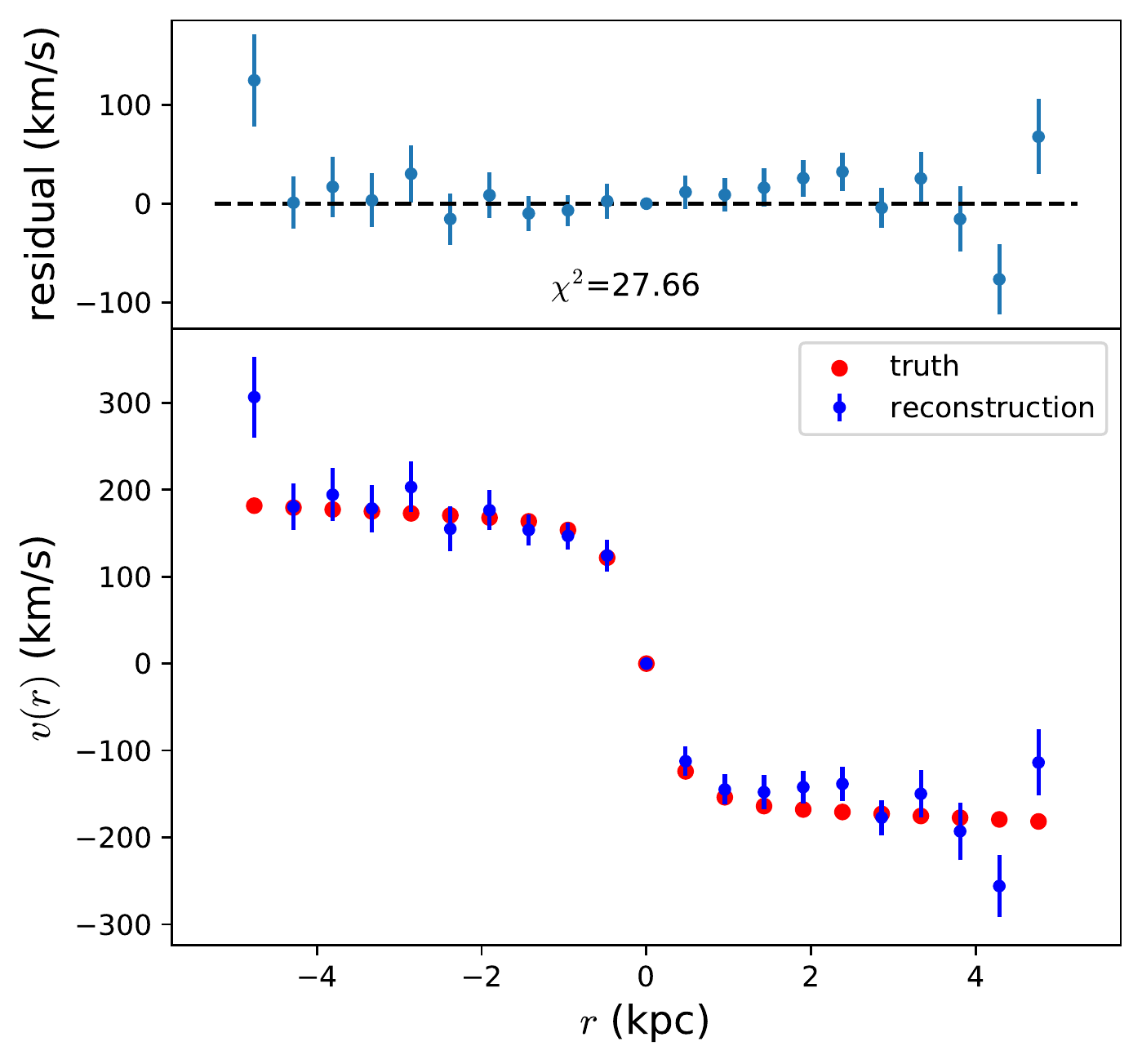}\label{fig:SN_test_abpVHN}}
\caption{Velocity reconstructions for spaxels with 
absorption line dominated spectra, 
with different noise levels. 
}
\label{fig:SN_test_abp}
\end{figure}

To provide more robust estimations for spectra with lower S/N, we use $\Delta=4.0$ \AA\ for analysing the medium and higher noise sets but keep $\Delta=1.5$ \AA\ for the low noise set.
The rotational velocities estimated for the four sets with different noise levels are compared with the truth 
(red dots) in the  panels of the figure~\ref{fig:SN_test_abp}. 
The method is again successful, even for such low 
S/N, though the 
uncertainties increase noticeably in the higher 
noise cases.

\begin{figure}
\centering
\includegraphics[width=0.485\textwidth]{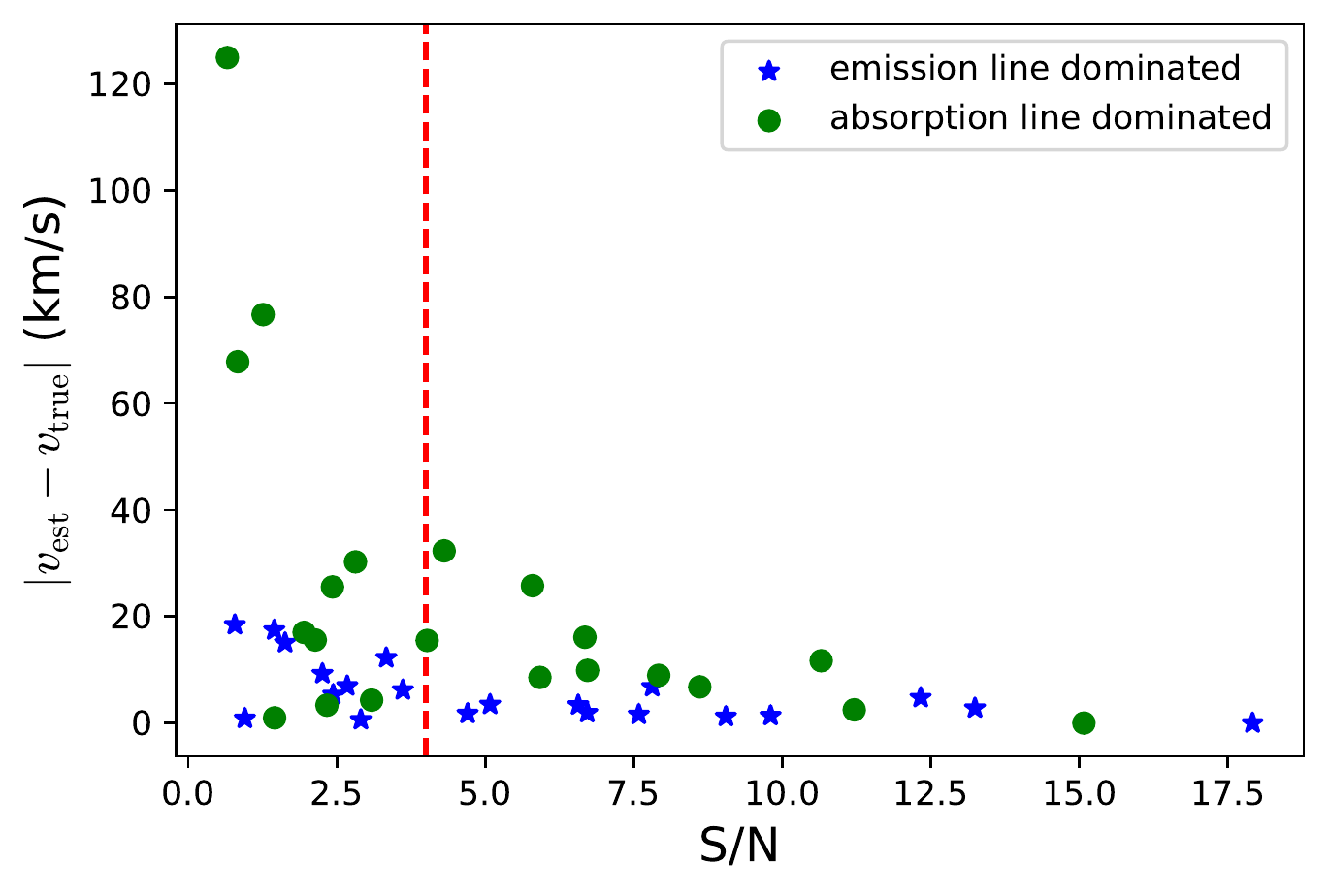}
\caption{The figure shows the absolute residuals as a function of S/N for the very high noise sets from the tests with the emission line dominated spectra (shown by blue stars) and absorption line dominated spectra (shown by green dots). The corresponding velocity estimations are already shown in figures \ref{fig:SN_test_emission_VHN} and \ref{fig:SN_test_abpVHN} with the uncertainties. The red vertical line represents S/N$=4$ below which the traditional template fitting approach struggles to estimate rotational velocities.}
\label{fig:VHnoise}
\end{figure}

Let us now focus on the results of the very high noise sets from the two cases, with emission line dominated spectra (test demonstrated in section \ref{sec:SN_test_emission}) and with absorption line dominated spectra (test demonstrated in section \ref{sec:SN_test_absp}). In figure \ref{fig:VHnoise}, we explicitly show the absolute values of the residuals of velocities (for spaxels) as a function of S/N for these two sets. The red vertical line represents S/N $=4$ below which the traditional template fitting approach struggles to estimate rotational velocities. Note that the velocity estimations, along with the uncertainties, have been shown in figures \ref{fig:SN_test_emission_VHN} and \ref{fig:SN_test_abpVHN} respectively for these two sets.

We find that our estimations continue to be reasonably accurate well beyond S/N $=4$ for both sets. For the case with emission line dominated spectra, we get residual $\lesssim 20$ km/s for this set when S/N $\sim 0.8$ only (due to the strong emission lines barely emerging from the noisy continua in the bottom panel of figure \ref{fig:SN_emission_spectra}). Even for the set with absorption line dominated spectra the results are accurate till S/N $\sim 1.3$, below which we may get quite large deviations from the true velocities but still the results are consistent with the truth within $2\sigma$ as evident from figure \ref{fig:SN_test_abpVHN}. 
 As demonstrated in appendix \ref{tab:ppxf_comp}, our results for the absorption lines dominated sets with four different noise levels are superior to the velocity estimations using the traditional fitting approach based on Penalized Pixel-Fitting (pPXF) \citep{Cappellari2004,2017MNRAS.466..798C} and MILES stellar spectral library \citep{Vazdekis2010}.

\section{Application to an observed MaNGA galaxy} \label{sec:data} 

Finally, we calculate the galaxy rotation curve for the MaNGA galaxy 7991--12701 itself. We estimate the velocity differences between spaxels along a major axis on the IFU hexagon, from spaxel ($37,12$) to spaxel ($37,64$)\footnote{We find maximal variation in velocity difference along this major axis. The emission line dominated spectra in some of the spaxels along this  axis have been shown in figures \ref{fig:demo_spectra} and \ref{fig:emission_spectra_7991-12701}.} for simplicity. The spectra have been observed with good signal to noise ratio along this axis; S/N $\sim 47$ for the spectrum in the central spaxel (37,37) and it falls down to $\sim 14$ at the outskirts.
We apply the same approach to the observed 
data as we had to simulated data.

\subsection{Velocity difference} 
\begin{figure}
\centering
\includegraphics[width=\textwidth]{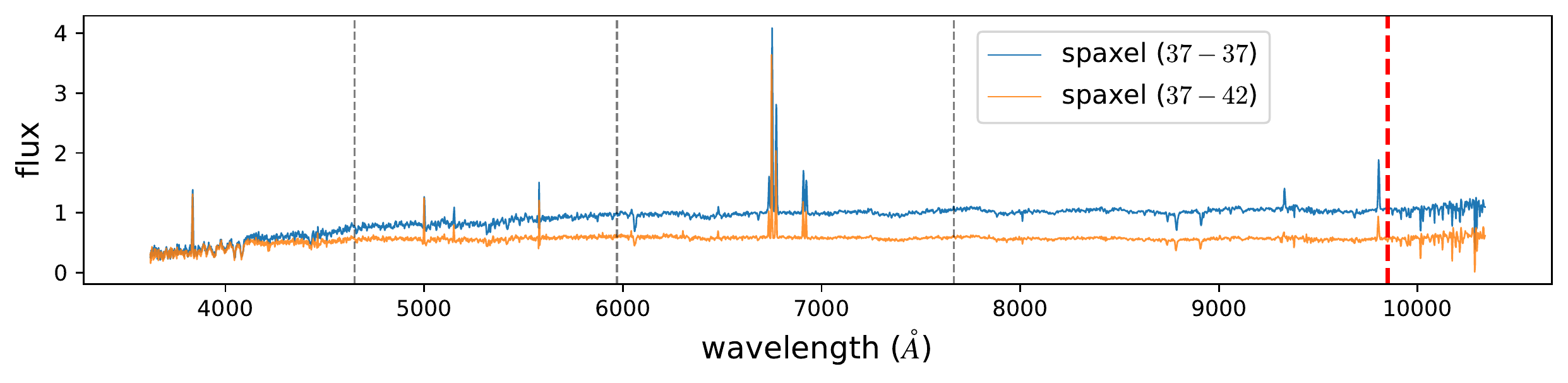}
\caption{Observed spectra from the spaxels (37,37) and (37,42) of the MaNGA galaxy 7991-12701 are dominated by emission lines (the flux is measured in the unit of [\flu]). We discard the data with $\lambda > 9850$ \AA, i.e.\ beyond the red vertical dashed line, for telluric contamination and  divide the rest of the data into 4 bins, shown by the vertical dashed lines. Note that for these observed spectra the wavelength separation increases with wavelength, leading to different bin widths.}
\label{fig:demo_spectra}
\end{figure}

An example of the actual spectra for a pair of spaxels, $A=(37,37)$ and $B=(37,42)$, is
plotted in 
figure~\ref{fig:demo_spectra}, both are seen to be emission line dominated with the former having a redder continuum.

\begin{figure}
\centering
\subfigure[$37,37 \bigotimes 37,42$]{
\includegraphics[width=0.485\textwidth]{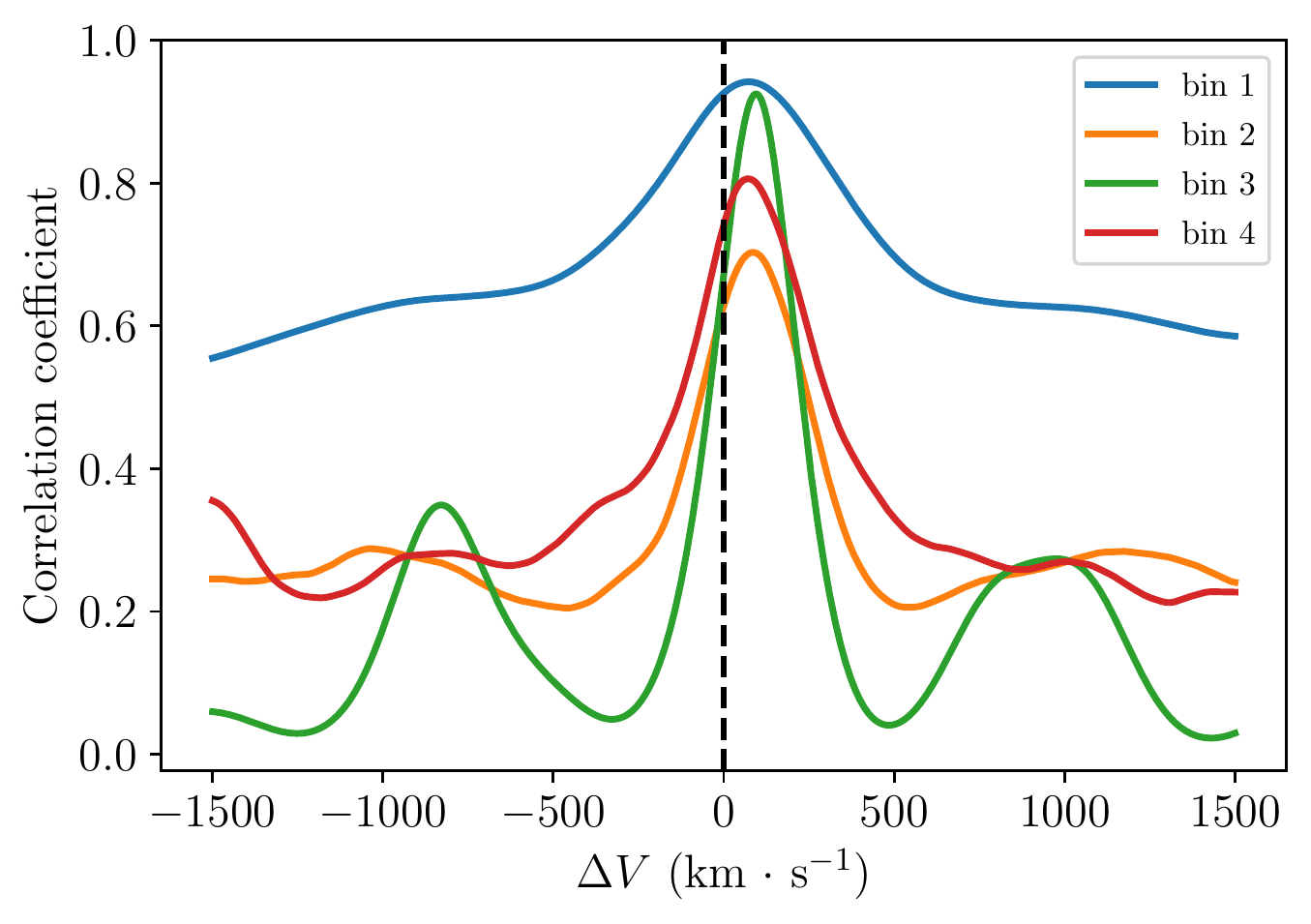}}
\subfigure[$37,42 \bigotimes 37,37$]{
\includegraphics[width=0.485\textwidth]{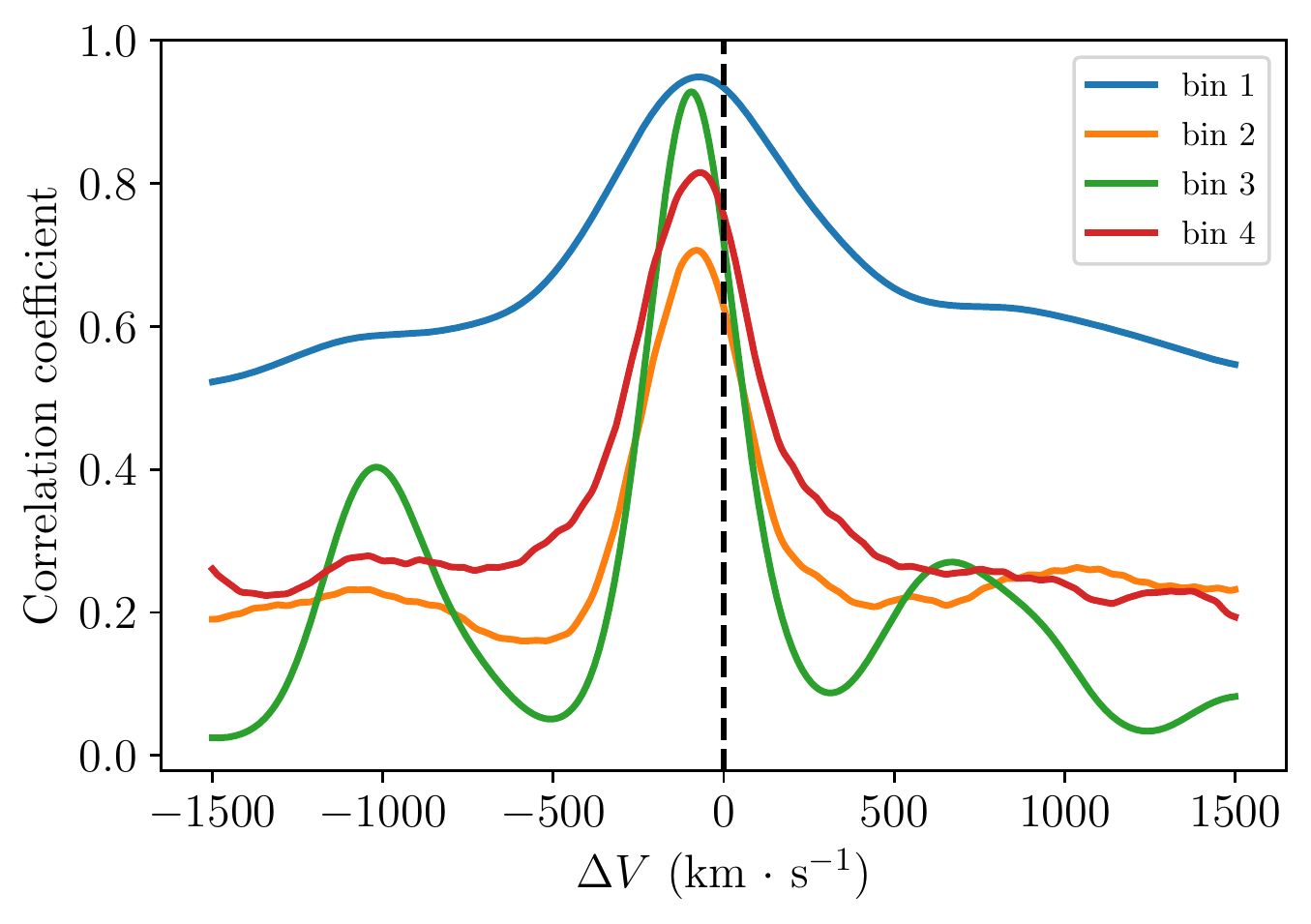}}
\caption{The correlation coefficient $r(\dv)$ is plotted as a function of velocity difference ($\dv$) between a pair of spaxels, $A=(37,37)$ and $B=(37,42)$ from the MaNGA galaxy 7991-12701. The left and right panel show $r_{A^sB}$ and $r_{B^s A}$ respectively for $4$ bins, calculated using \eqref{eq:rab}. All the bins have maxima in similar $\dv$ in either panel and the peak 
values are the negative of each other, as expected. 
}
\label{fig:demo}
\end{figure}

In figure~\ref{fig:demo}, we show the correlation coefficient as a function of the velocity difference, $r(\dv$), between the pair of spaxels. The left and right panels show $r_{A^sB}$ and $r_{B^s A}$ respectively for $4$ bins. 
For this illustration we choose $\D=1.5$ \AA\ and $\nit=10$. In either panel we find that $r$ for different bins exhibits global maxima at similar values of $\dv$ (and $\dv_{BA} \approx - \dv_{AB}$). 
More importantly, although bin 4 does not contain any strong feature, as evident from 
figure~\ref{fig:demo_spectra}, due to small spectral features $r$ shows a prominent peak at a consistent $\dv$ in both panels of 
figure~\ref{fig:demo} for this bin. This illustrates the key benefit of this method that it does not rely on strong identifiable features; rather makes use of the whole spectra. 
Thus we have $4 \times 2=8$ `good' estimations of $\dv$ from both the panels. 
We estimate the velocity difference between these two spaxels by taking the mean and standard deviation of these eight individual estimations, 
obtaining $\dv_{AB} \approx 80.29 \pm 9.82$ \kmps.

In table~\ref{table:manga_7991-12701_32_42}, we present the estimations of $\dv$ for the same pair of spaxels as in figure~\ref{fig:demo} but for a number of different $\D$, $\nit$, and number of bins. The $\dv$ estimations corresponding to different values of these parameters are consistent with each other that demonstrates the robustness of this approach. Since the spectra in the spaxels we consider here have good S/N ($\gtrsim 14$), we adopt $\D=1.5$, $\nit=10$ and 4 bins in the analysis of this MaNGA galaxy, same as what we used for the simulations in section \ref{sec:validsim}.
 
Note the value of $\dv$ obtained from the stellar velocity map by the Marvin team is $85.89 \pm 3.34$ km/s, consistent with our estimations. 
There is nothing to say what the true answer is. 
However, the Marvin results come from their full 2D analysis, giving extra stability to the results and enforcing the full triangle equality   (that the vector sum of velocity differences among a triangle of spaxels is zero). Having demonstrated our correlation method here, we plan to apply it to the 2D data in the follow-up paper.

\begin{table*}
\begin{tabular}{|c|c|c|c|c|}
 \hline
 \multicolumn{1}{|c|}{\multirow{2}{*}{$\Delta$ (in \AA)}} & \multicolumn{2}{|c|}{4 bins}& \multicolumn{2}{|c|}{8 bins}\\\cline{2-5}
 & $\nit=10$ & $\nit=20$ &  $\nit=10$ & $\nit=20$\\
 \hline
 $1.5$  & $ 80.3 \pm 9.9 $ & $ 78.8 \pm 10.1 $ & $  83.5 \pm 10.1 $ & $ 83.2 \pm 10.6 $  \\
 \hline
 
  $2.0$  & $ 84.0 \pm 9.5 $ & $ 82.5 \pm 9.9$ & $ 82.0 \pm 10.5 $ & $  84.0 \pm 9.5$  \\
 \hline
 
  $3.0$  & $ 85.1 \pm 8.4 $ & $ 85.7 \pm 8.5 $ & $ 84.9 \pm 8.9$ & $ 85.2 \pm 8.4$  \\
 \hline
 
  $4.0$  & $ 82.8 \pm 9.2 $ & $ 83.7 \pm 8.6 $ & $ 84.7 \pm 10.6 $ & $ 85.4 \pm 10.1 $  \\
 \hline
 
 \hline

 \hline
\end{tabular}
\caption{Velocity difference between MaNGA 7991-12701 spaxels (37,37) and (37,42) for different values of smoothing scale $\Delta$, 
number of iterations $\nit$, and number of wavelength bins considered. The value quoted in Marvin DR15 
between these two spaxels is 
$85.89 \pm 3.34$ km/s . 
}
\label{table:manga_7991-12701_32_42}
\end{table*}

We list several more velocity differences 
from spaxel pairs of  
MaNGA galaxy 7991--12701 in 
table~\ref{tab:7991_12701_spaxels}, 
along with the $\dv$ values obtained from the 
stellar velocity map from Marvin.
 
Our estimations are somewhat consistent with the Marvin values but tend 
to be slightly higher than that from the Marvin stellar map. However, note that two estimations based on two different approaches have uncertainties of similar order of magnitude. 
In the simulation studies our method estimated $\dv$ quite accurately, without 
a bias or underestimation of uncertainties, so it is not clear what 
the true answer is. The full 2D data 
analysis in our follow-up paper, now 
that we have demonstrated the method, 
will give a more parallel comparison 
to Marvin.

\begin{table}
\centering
 \begin{tabular}{||c|c|c||} 
 \hline
 Pair of spaxels (A,B) & Our estimation of $\Delta V_{AB}$ in km/s & $\Delta V_{AB}$ in km/s from Marvin\\ [0.5ex] 
 \hline\hline
 ($37$,$12$) , ($37$,$36$) & $137.45 \pm 3.65$ & $130.45 \pm 6.39$ \\
 \hline
 
 ($37$,$16$) , ($37$,$37$)  & $153.84 \pm 4.31$ & $152.60 \pm 5.95$ \\
 \hline
 ($37$,$20$) , ($37$,$41$) & $223.99 \pm 4.29$ & $216.36 \pm 5.33$ \\ 
 \hline
 ($37$,$22$) , ($37$,$38$) & $167.38 \pm 4.91$ & $171.15 \pm 5.16$ \\
 \hline
 ($37,26$) , ($37,32$) & $ 32.43 \pm 3.61 $  & $ 39.54 \pm 4.78$    \\ 
 \hline
 
  ($37,26$) , ($37,37$) & $  135.69 \pm 4.62$  & $ 128.94 \pm 4.45$   \\ 
 \hline
 
 ($37$,$28$) , ($37$,$36$) & $100.48 \pm 2.72$ & $98.94 \pm 4.15$ \\ 
 \hline

 ($37,32$) , ($37,37$) & $ 98.62 \pm 5.19 $  & $ 89.40 \pm 3.37$   \\ 
 \hline
 
 ($37,37$) , ($37,42$) & $80.29 \pm 9.82  $  & $85.89 \pm 3.34$   \\ 
 \hline
 
 ($37,37$) , ($37,48$) & $ 143.79 \pm 6.34 $  & $131.70  \pm 3.65$   \\ 
 \hline
 
 ($37$,$37$) , ($37$,$58$) & $171.08 \pm 2.81$ & $166.68 \pm 5.19$ \\
 \hline
 
 ($37$,$38$) , ($37$,$60$) & $155.67 \pm 5.25$ & $151.91 \pm 5.85$ \\
 \hline
 
 ($37$,$41$) , ($37$,$64$) & $90.19 \pm 4.29$ & $86.77 \pm 5.42$ \\ 
 \hline

 \hline

 \hline
\end{tabular}
\caption{We compare our estimation of the velocity difference $\Delta V_{AB}$ between various pairs of  spaxels $A$ and $B$ with that from Marvin DR15. Here the smoothing scale $\Delta=1.5$ \AA, $\nit=10$, and we divided the spectra into $4$ wavelength bins.}
\label{tab:7991_12701_spaxels}
\end{table}

\subsection{Constructing the galaxy rotation curve} 
\label{sec:grcmanga} 

Now we complete constructing the galaxy rotation curve for this MaNGA galaxy, using the spectra in the spaxels along the major axis on the IFU hexagon, from spaxel ($37,12$) to spaxel ($37,64$).
Calculating $\dv$ for all the pairs of spaxels along this major axis would be more computationally expensive than warranted now for our purpose of simple demonstration. Thus we choose 
the 11 central spaxels as 
anchor spaxels (i.e.\ the $j$ spaxels of section~\ref{sec:combine}),  
and add 23 spaxels more 
over the range where the data quality is good 
to the $i$ spaxel set, 
giving a total of 34 spaxels in the HMC analysis.

\begin{figure}
\centering
\includegraphics[width=0.6\textwidth]{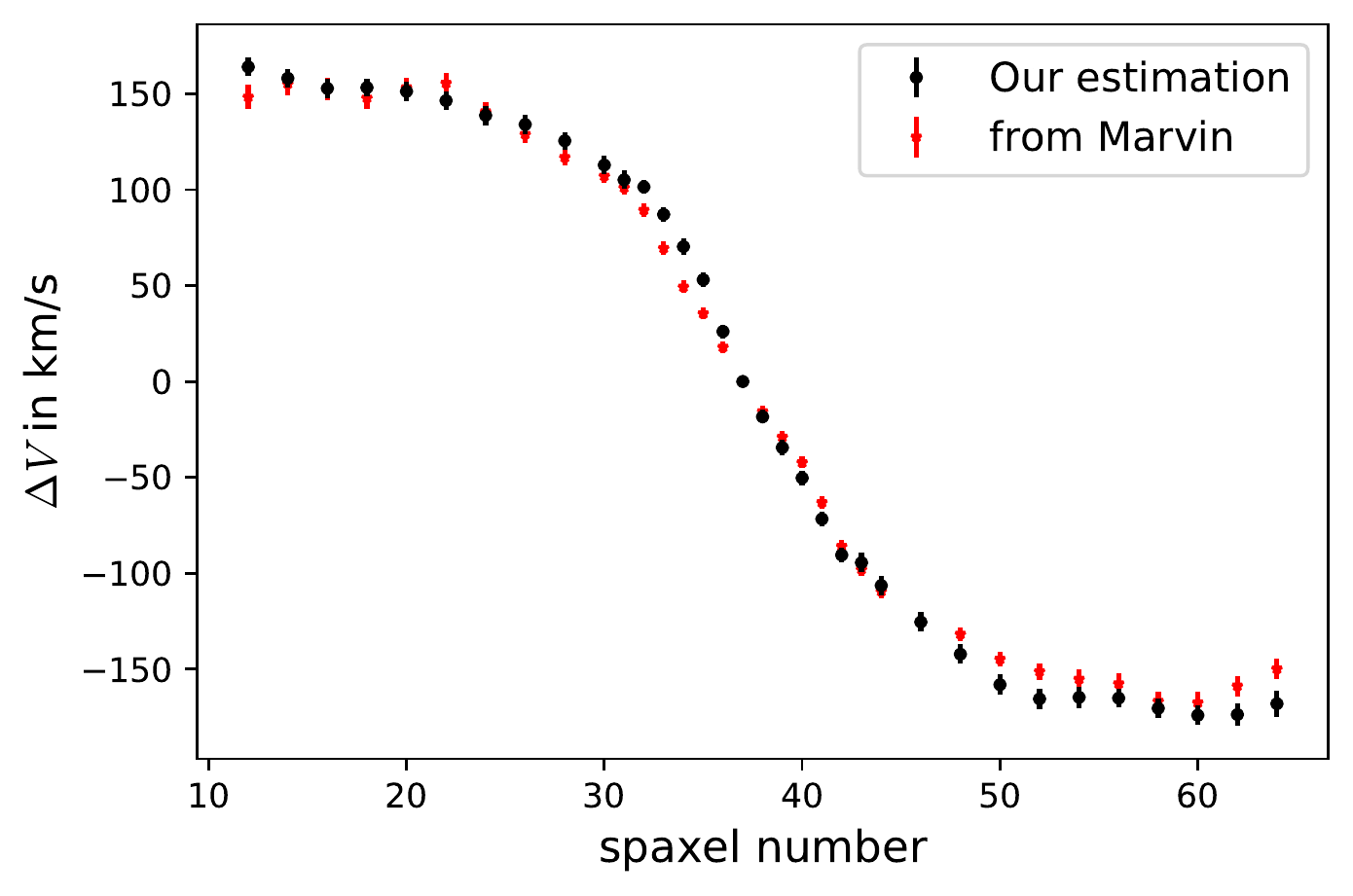}
\caption{Our estimations (black dots with error bars) of velocity along the line of sight of the spaxels on the chosen axis with respect to the central spaxel $(37,37)$ of the MaNGA galaxy 7991--12701. The uncertainties 
are so small that the errorbars are not visible for most of the spaxels.
We compare to the velocities given in the Marvin database (red dots with error bars).
Note that S/N $\approx 47.5$ at the centre and gradually falls down to $\sim 14$ at outskirts.}
\label{fig:manga_7991_12701}
\end{figure}

In figure \ref{fig:manga_7991_12701} we compare our estimated velocities (relative to the central spaxel) with that from the Marvin for the selected spaxels.
As noted in the previous section, we find reasonable agreement between Marvin results and our estimations, with some trend toward larger velocities. Again we find that the uncertainties in our estimations are comparable to that from Marvin across the spaxels; the errorbars are too small to be visible for most of the central spaxels. 
Uncertainties in the outer disk region are larger, but we do not find a decrease in velocity as Marvin does. Our forthcoming  
2D analysis may shed further light.

\section{Conclusions and discussion}
\label{sec:conclusion}

Galaxy rotation curves, or the internal dynamics in general, provide one of the primary lines of evidence 
for the existence of dark matter. Besides merely indicating the 
presence of dark matter, one can map the velocity-radius relation  
$v(r)$ directly to the mass distribution with radius. 
Traditionally, rotational velocities are obtained by first fitting a 
template to the spectra of different regions, e.g.\ spaxels along a galaxy 
major axis, and then calculating the Doppler shift between spaxels. The Doppler shifts are then translated into the velocity differences between the pairs of spaxels.
This works, but has dependence on the template accuracy for that 
particular galaxy type and spectral noise properties.  

In this article, we present a novel and template-free method to calculate the galaxy rotation curves based on cross-correlation between the spectra in the IFU data, with each member of the comparison pair alternately smoothed to avoid spurious features from noise. 
We demonstrate that one can achieve accurate and precise velocity difference measurements with this method. 
We then globally optimize the array of velocity difference through 
Hamiltonian Monte Carlo to construct the full galaxy rotation curves for galaxies, with spectra simulated in a variety of observational conditions. 

The test conditions include emission line dominated spectra with red/blue continuum, absorption line dominated spectra, 
switching between different types of spectra, injection of various levels of noise, etc.
The method appears to be promising and the results are precise with small uncertainties in most cases, in particular for spaxels with S/N $\gtrsim 4.0$. Remarkably, even for very noisy spectra with S/N $\sim 1$ we are able to recover the true rotational velocities with reasonable accuracy. 

After these validation tests we apply the method to data from the observed MaNGA galaxy 7991--12701. Our estimations of velocity differences (between different pairs of spaxels) appear to be slightly higher than what was obtained from the Marvin velocity field, but for most cases they are consistent with each other. Comparing the 1D 
rotation curve (i.e.\ 
the velocities of the spaxels with respect to the central one) 
between our method and Marvin, 
we find that the results are mostly consistent with each other 
(figure~\ref{fig:manga_7991_12701}). However, for the outer disk our estimated rotation curve does not show the slight reduction in  
rotational velocities seen in Marvin. The results are 
sufficiently promising to pursue further. 

One of the chief advantages of this method is that it does not rely on strong identifiable features in the spectra. Rather, cross-correlation 
can utilize many small features in the spectra, and in conjunction with smoothing, remain fairly insensitive to noise. We have found 
good results even for quite noisy spectra. 

There are several aspects to pursue further, going beyond the 
present proof of concept. We will enhance the pipeline to analyse the whole 2D IFU data of a number of MaNGA galaxies, not just a 1D 
slice on a diameter/major axis. This should also add further 
stability to the results through the triangle equality within 
our HMC analysis. We adopted here a simplified algorithm to construct the galaxy rotation curve for illustrating our results; the general 
conversion from a well measured velocity field to constructing the 
galaxy rotation curve has additional elements. 
We also plan to focus on low surface-brightness galaxies where the traditional template fitting approach struggles, especially at the outskirts. Further investigation will show the degree to which 
this new method can improve upon, or 
nicely complement, traditional template fitting approaches.

\section*{Acknowledgement}
The authors acknowledge that the high performance computing facility at the Korea Institute of Science and Technology Information (KISTI), assignment no. KSC-2020-CRE-0153, has been used in this project. SB and AS thank Adarsh Ranjan for crucial helps and Alex G. Kim for useful discussions at different phases of the project. S.B. also thanks Brian Cherinka and Maria Argudo-Fernández for useful discussions and explanations regarding the MaNGA data.
YSA, EL, and KY were supported in part by the Energetic Cosmos Laboratory and EL by the U.S.\ Department of Energy, Office of Science, Office of High Energy Physics, under contract no.\ DE-AC02-05CH11231. KY thanks Yessenov Foundation for funding his stay in South Korea. YSA and KY also thank KASI for hospitality during the early work.

\section*{Data availability}

The data underlying this article will be shared on reasonable request
to the corresponding author.

\appendix

\section{Comparing with Penalized Pixel-Fitting (pPXF)}
In this section we compare our results with that from the traditional fitting approach based on Penalized Pixel-Fitting (pPXF) \citep{Cappellari2004,2017MNRAS.466..798C}, using the MILES templates from \citet{Vazdekis2010}. Since obtaining the fitting results for all the cases is beyond the scope of this article, we restrict ourselves to the data sets with absorption line dominated spectra. In particular, we consider the four data sets with different noise levels as described in section \ref{sec:SN_test_absp}. Our results for these tests have been presented in figure \ref{fig:SN_test_abp} already. 

\begin{figure}
\centering
\subfigure[low noise set (S/N $\gtrsim 37.39$)]{\label{fig:SN_test_pPXF_absorp_low}
\includegraphics[width=0.49\textwidth]{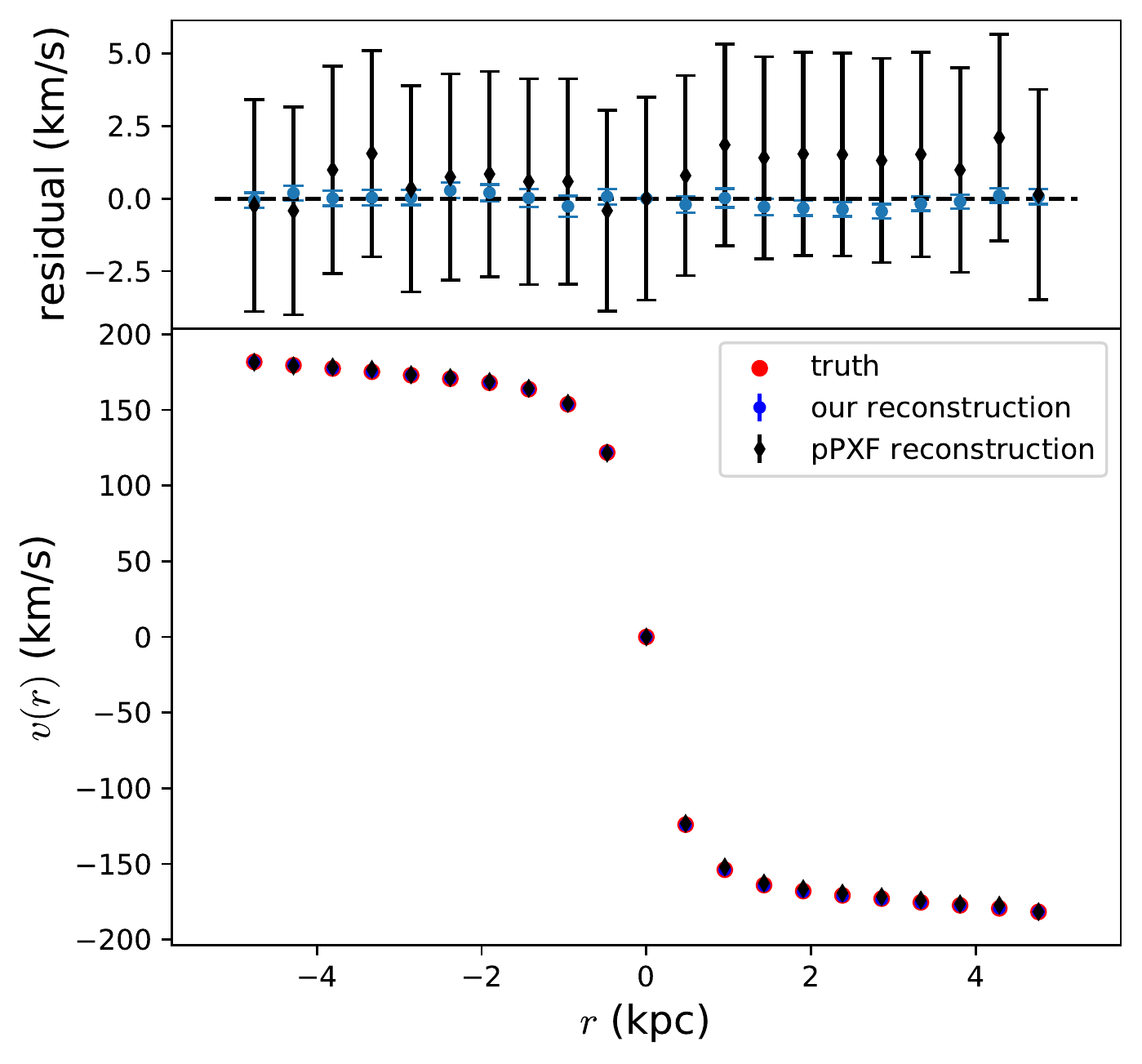}\hspace*{-2mm}}
\subfigure[medium noise set (S/N $\gtrsim 9.95$)]{
\includegraphics[width=0.49\textwidth]{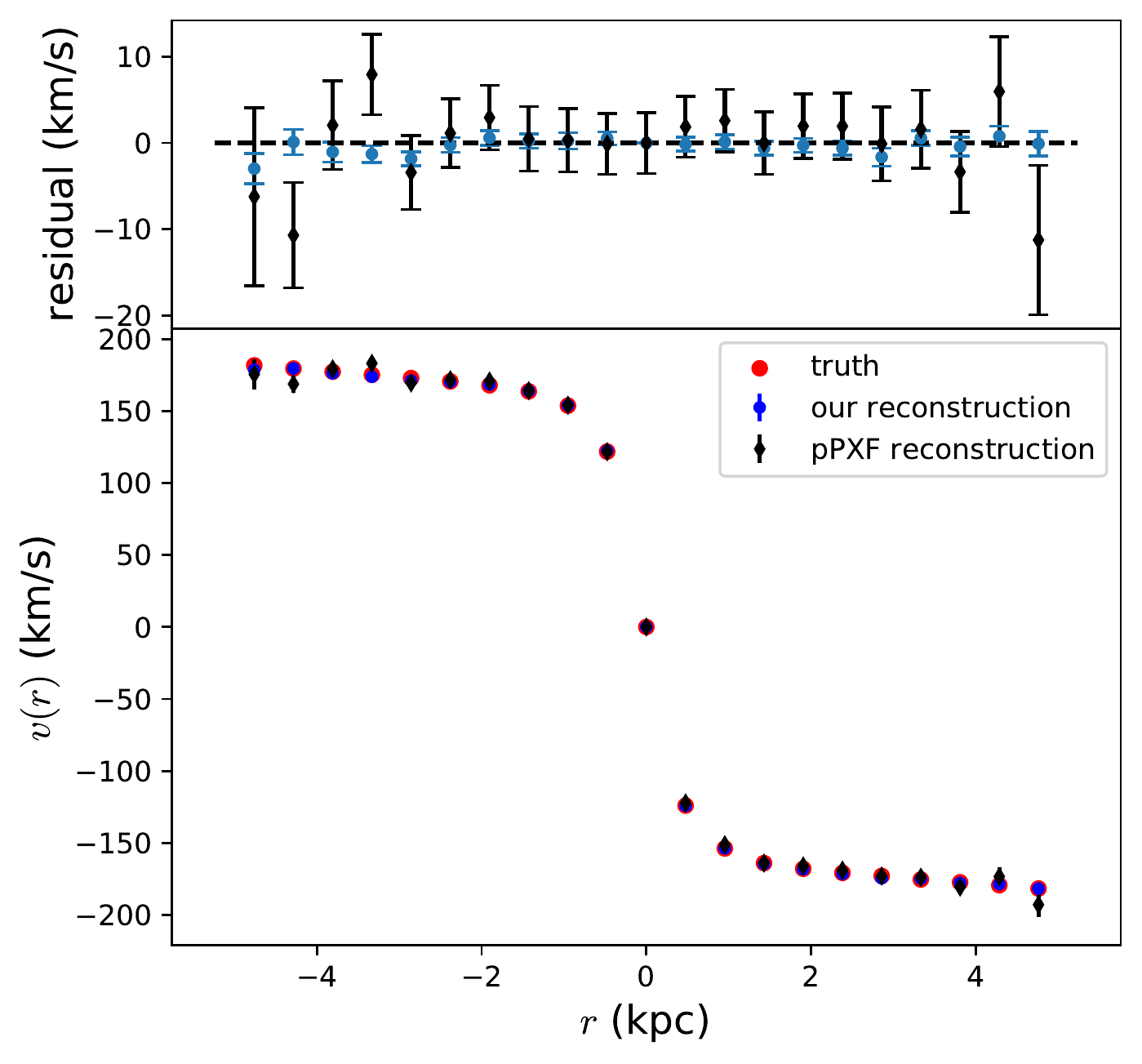}}\\
\subfigure[high noise set (S/N $\gtrsim 2.05$)]{
\includegraphics[width=0.49\textwidth]{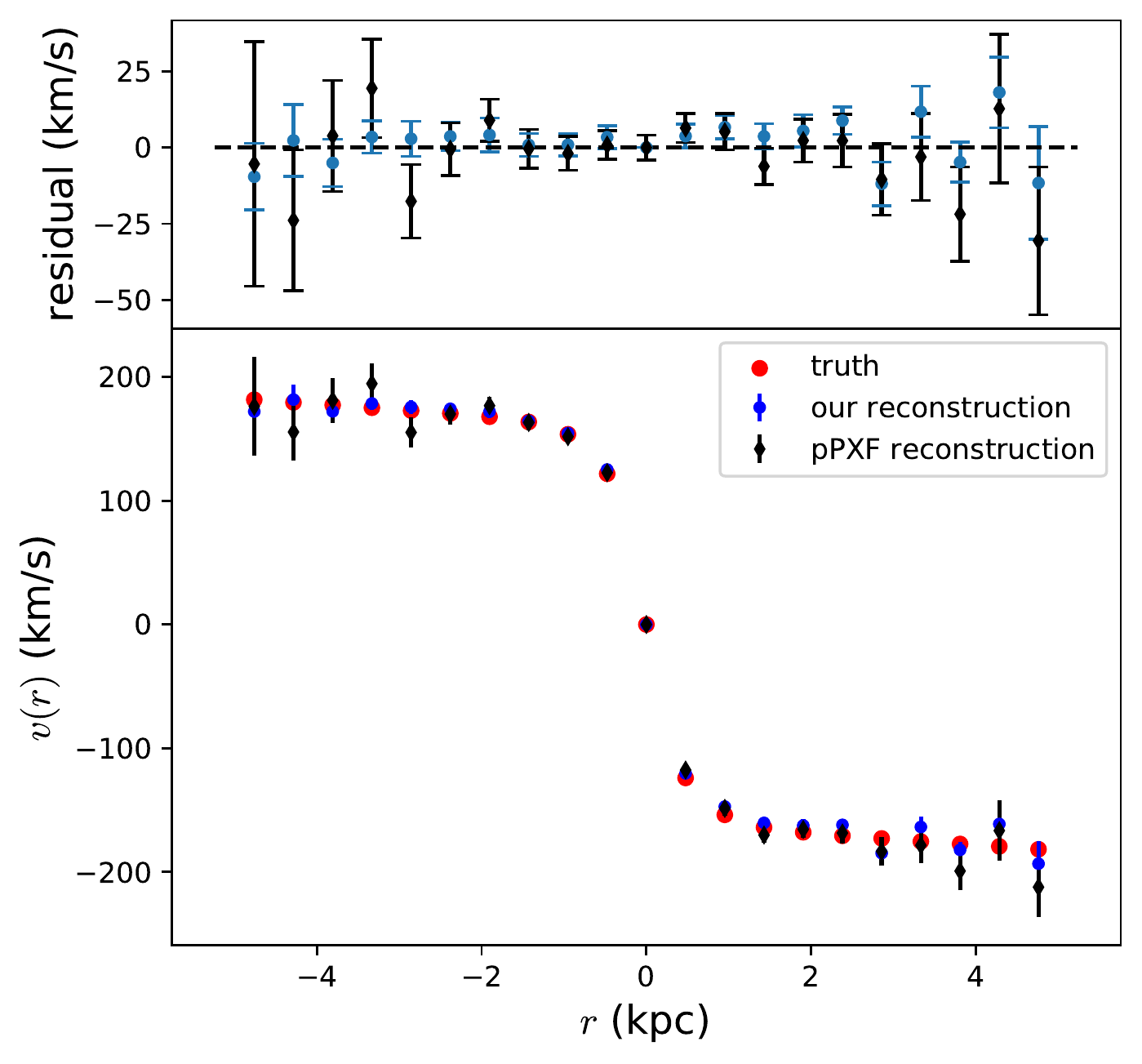}\hspace*{-2mm}}
\subfigure[very high noise set (S/N $\gtrsim 0.66$)]{
\includegraphics[width=0.49\textwidth]{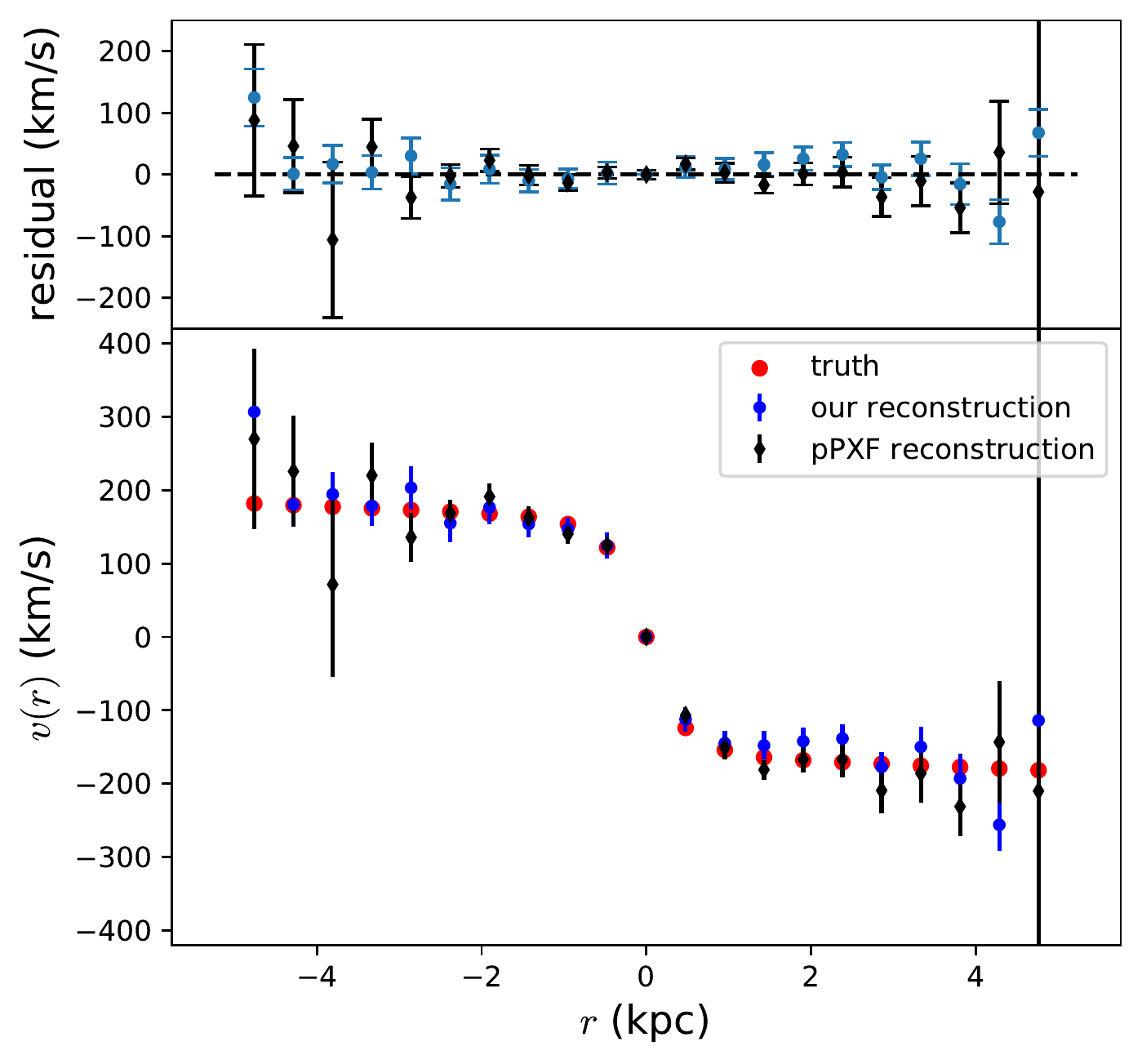}}
\caption{Comparing the velocity reconstructions from our approach (blue) and pPXF (black) for the 
absorption line dominated spectra data sets with different noise levels (this test and our results are explained in section \protect\ref{sec:SN_test_absp} in more details). Note that the errorbar from pPXF for the rightmost spaxel in the right-bottom panel is unrealistically high ($\sim 9000$\kmps, truncated by hand in the plot) which indicates that the automated pPXF is unstable for such low S/N cases. Also note that pPXF can overestimates the uncertainty on the fit velocity as vividly evident in the low-noise case presented in the top-left panel. 
}
\label{fig:SN_test_abp_ppxf}
\end{figure}

As evident from figure \ref{fig:SN_test_abp_ppxf}, the velocities estimated from pPXF are in good agreement with the truths for all four sets, especially for lower noise sets (higher S/N cases) presented in the top panels. However, for the higher noise sets, pPXF results becomes less accurate which can be seen clearly in residual plots presented in the top subplots of the bottom two panels. Nevertheless, our estimations are superior than that from pPXF, as expected, in all four cases. The uncertainties from pPXF are much larger than ours for all the sets, although we find that pPXF overestimates the uncertainties for many spaxels (especially for the low noise level cases in the top panels \protect\footnote{When we apply pPXF on the base MaNGA spectrum that was used for the simulation and has similar noise level as the low-noise case, we obtain velocity errors of the same order ($\sim 3$ \kmps) as in the top-left panel of figure \ref{fig:SN_test_abp_ppxf}.}). 
Also, we notice that the automated pPXF is not robust in the very high noise cases as it suffers from occasional failings as well as produces unrealistically huge errors sometimes (bottom-right panel).

We compare the accuracy of the velocities  estimated from the two approaches in terms of the following quantities
\beq
\epsilon_r=\frac{1}{N_{\rm spax}}\sum_i \left|v_{\rm est}-v_{\rm true} \right|~,~~\epsilon_p=\frac{1}{N_{\rm spax}}\sum_i \left|\frac{v_{\rm est}-v_{\rm true}}{v_{\rm true}}\right|~,~~b=\frac{1}{N_{\rm spax}}\sum_i \left(v_{\rm est}-v_{\rm true}\right)\;,
\eeq
which are the average residual, the average percentage error and the bias respectively.
Table \ref{tab:ppxf_comp} compares the results of these two approaches for the four sets in terms of $\epsilon_r$, $\epsilon_p$ and bias ($b$). While both method shows insignificant bias, the precision (in terms of $\epsilon_r$, $\epsilon_p$) of the spaxel cross-correlation approach is found be always better than that of pPXF.

\begin{table}
\centering
 \begin{tabular}{|c|c|c|c|c|c|c|} 
 \hline
  \multicolumn{1}{|c|}{\multirow{2}{*}{Set}} & \multicolumn{2}{|c|}{Average residual ($\epsilon_r$) in \kmps}& \multicolumn{2}{|c|}{Percentage error ($\epsilon_p$)}& \multicolumn{2}{|c|}{Bias ($b$) in \kmps}\\\cline{2-7}
 & This work & pPXF & This work & pPXF & This work & pPXF  \\[0.5ex]
 \hline
Low noise & $ 0.16$ & $ 0.99 $ & $ 0.1\%$ & $  0.6\%$ & $ 1.37 \cdot 10^{-3}  $ & $ -2.66\cdot 10^{-3}$\\
 \hline
 
Medium noise & $ 0.72$ & $ 3.29 $ & $ 0.42\%$ & $  1.89\%$ & $ -1.62 \cdot 10^{-3}  $ & $-2.21 \cdot 10^{-3} $\\
 \hline
 
High noise & $ 6.14$ & $ 9.20 $ & $ 3.62\%$ & $   5.33\%$ & $ 1.35 \cdot 10^{-2}  $ & $6.68 \cdot 10^{-3} $\\
 \hline

Very high noise & $25.24 $ & $ 28.60 $ & $ 14.52\%$ & $  16.50\%$ & $2.87 \cdot 10^{-2}   $ & $3.42 \cdot 10^{-2} $\\
 \hline
\end{tabular}
\caption{We compare the results of the two approaches -- spaxel cross-correlation vs pPXF -- for the four sets with different noise levels in terms of average residual, average percentage error and bias.}
\label{tab:ppxf_comp}
\end{table}

This exercise thus demonstrates that our approach based on spaxel cross-correlation provides more precise estimation of velocities as it use the information from all parts of the spectra. However, this comes at the expense of moderate amount associated computation cost. pPXF takes roughly $0.5$ seconds on a 32-thread CPU to fit one MaNGA-like spectrum, whereas it takes $\sim 30$ sec to obtain the velocity difference between two such spectra using spaxel cross-correlation on the same computer. Nevertheless, the robustness for low S/N spectra and higher precision in our approach possibly justify the tread off in some particular use cases.

\bibliographystyle{mnras}
\bibliography{grc_cc}

\end{document}